\documentclass[sn-mathphys-num]{sn-jnl}

\usepackage{graphicx}%
\usepackage{multirow}%
\usepackage{amsmath,amssymb,amsfonts}%
\usepackage{amsthm}%
\usepackage{mathrsfs}%
\usepackage{xcolor}%
\usepackage{textcomp}%
\usepackage{manyfoot}%
\usepackage{booktabs}%
\usepackage{soul}
\usepackage{mathtools}
\usepackage{bm}
\usepackage[normalem]{ulem}

\usepackage{xr}

\newcommand{\didv}{{d}\textit{I}/{d}\textit{V} }

\newcommand{\bra}[1]{\left\langle#1\right|}
\newcommand{\ket}[1]{\left|#1\right\rangle}
\newcommand{\bracket}[2]{\big\langle#1 \bigm| #2\big\rangle}

\newcommand{\up}{\uparrow}
\newcommand{\dn}{\downarrow}

\newcommand{\Up}{\Uparrow}
\newcommand{\Dn}{\Downarrow}

\newcommand{\Rt}{\Rightarrow}

\newcommand{\Egs}{E_{\rm gs}^{(0)}}

\newcommand{\JK}[1]{J_{#1}}
\newcommand{\JKT}[1]{\tilde{J}_{#1}}
\newcommand{\TK}{T_{\rm K}}
\newcommand{\HK}{\mathcal{H}_{\rm K}}
\newcommand{\VK}{\mathcal{V}_{\rm K}}
\newcommand{\HKT}{\tilde{\mathcal{H}}_{\rm K}}

\newcommand{\CJ}{\mathcal{J}}

\newcommand{\eV}{{\rm eV}}
\newcommand{\meV}{{\rm meV}}

\begin{document}

\title[Article Title]{Coexistence of Ferromagnetic and Overscreened Kondo Effects in a Single Molecule}


\author*[1,2]{\fnm{Elia} \sur{Turco}}
\email{e.turco@tudelft.nl}

\author[1]{\fnm{Nils} \sur{Krane}}

\author[3]{\fnm{Hongyan} \sur{Chen}}

\author[3]{\fnm{Simon} \sur{Gerber}}

\author[3]{\fnm{Wulf} \sur{Wulfhekel}}

\author[1,4]{\fnm{Roman} \sur{Fasel}}

\author[1]{\fnm{Pascal} \sur{Ruffieux}}

\author*[5]{\fnm{David} \sur{Jacob}}
\email{david.jacob@ua.es}

\affil[1]{\small\orgdiv{nanotech@surfaces Laboratory}, \orgname{Empa -- Swiss Federal Laboratories for Materials Science and Technology},
  \orgaddress{\postcode{8600} \city{D\"ubendorf} \country{Switzerland}}}

\affil[2]{\small Current address: QuTech and Kavli Institute of Nanoscience, Delft University of Technology, 2600 GA Delft, The Netherlands}

\affil[3]{\small\orgdiv{Physikalisches Institut}, \orgname{Karlsruhe Institute of Technology},
   \orgaddress{\postcode{76131} \city{Karlsruhe}, \country{Germany}}}

\affil[4]{\small\orgdiv{Department of Chemistry, Biochemistry and Pharmaceutical Sciences}, \orgname{University of Bern},
   \orgaddress{\postcode{3012} \city{Bern}, \country{Switzerland}}}

\affil[5]{\small\orgdiv{Departamento de F\'isica}, \orgname{Universidad de Alicante},
   \orgaddress{Campus de San Vicente del Raspeig, \postcode{03690} \city{Alicante}, \country{Spain}}}

\abstract{
The quest for quantum ground states beyond the Fermi-liquid paradigm remains a central challenge in many-body physics. The ferromagnetic Kondo effect is an intriguing variant of the Kondo effect in which an asymptotically free spin gives rise to singular Fermi-liquid behavior. Despite its theoretical importance, this regime has eluded experimental observation owing to subtle spectroscopic signatures, small energy scales, and symmetry constraints in nanostructures.
Here, we demonstrate the coexistence of ferromagnetic and overscreened Kondo effects within a molecular spin system---a triangulene dimer comprising spin-1 and spin-1/2 units adsorbed on a metal surface. Low-temperature scanning tunneling spectroscopy reveals signatures of singular Fermi-liquid behavior, supported by many-body calculations. The molecular design provides control over spin configuration and coupling asymmetry, allowing distinct regimes within the same platform. Our results establish a strategy for realizing non-Fermi-liquid physics at the atomic scale and demonstrate that ferromagnetic Kondo behavior can be deliberately engineered in molecular systems.
}


\maketitle

\section*{Introduction}

The Kondo effect is a paradigmatic many-body phenomenon in condensed matter physics, arising from the interaction between a localized spin and conduction electrons in a metal. Classic realizations include magnetic atoms or molecules embedded in bulk metals or adsorbed onto metallic surfaces. While extensively studied for its fundamental relevance, the Kondo effect has found limited technological application, serving mainly as a spectroscopic probe of molecular magnetism in scanning tunneling microscopy~\cite{Madhavan:Science:1998,Li:PRL:1998,Manoharan:Nature:2000,Nagaoka:PRL:2002,Turco:PRR:2024}. In nanoscale magnetic systems, electronic coupling to metallic contacts underpins transport-based spin detection. Strong coupling, however, generates Kondo correlations that entangle the impurity with the leads, so that the local moment is no longer a well-defined low-energy degree of freedom. This fundamentally complicates spin manipulation and readout schemes predicated on isolated-spin physics. 

In certain multichannel or higher-spin settings screening can become frustrated or suppressed, preventing complete Kondo screening. This can give rise to exotic regimes such as the overscreened Kondo and ferromagnetic Kondo effects~\cite{Nozieres:JP:1980,Koller:PRB:2005}. Both evade complete screening but fall into distinct universality classes. The overscreened Kondo flows to a non-Fermi-liquid fixed point, leaving a fractional residual entropy that reflects fractionalized boundary excitations~\cite{Affleck:PRL:1991}. The ferromagnetic Kondo effect instead realizes a singular Fermi liquid in which the impurity retains an asymptotically free spin-1/2 leading to logarithmic corrections in thermodynamic and transport properties~\cite{Koller:PRB:2005,Baruselli:PRL:2013}. Beyond their conceptual interest, such exotic Kondo phenomena have been proposed as routes to non-Abelian boundary excitations---such as anyons and Majorana-like modes---with potential applications in topological quantum computation~\cite{Emery:PRB:1992,Potok:Nature:2007,Keller:Nature:2015,Iftikhar:Nature:2015,Lopes:PRB:2020}.

Despite their theoretical appeal, overscreened and ferromagnetic Kondo states remain experimentally difficult to access. The overscreened Kondo requires a finely tuned multichannel geometry where the number of screening channels $K$ exceeds twice the impurity spin $S$---a condition rarely realized in conventional nanostructures~\cite{Roch:PRL:2009}. Carefully engineered quantum dot devices have recently demonstrated overscreened Kondo behavior by achieving the necessary channel symmetry~\cite{Potok:Nature:2007,Keller:Nature:2015,Iftikhar:Nature:2015}. In contrast, ferromagnetic Kondo has so far remained elusive: although proposed in triple quantum dot architectures~\cite{Mitchell:PRB:2009,Baruselli:PRL:2013}
 and coupled atomic spins~\cite{ternes_probing_2017}, it has not been realized experimentally, likely due to unavoidable symmetry breaking and channel anisotropies in exchange-coupled nanostructures, which destabilize the ferromagnetic Kondo fixed point.

Here we show that a molecular nanographene dimer consisting of a spin-1 and a spin-1/2 triangulene adsorbed on Au(111) provides a natural platform where both regimes coexist.
Low-temperature scanning tunneling spectroscopy (STS) reveals characteristic signatures of singular Fermi-liquid behavior over the spin-1/2 unit and of non-Fermi liquid behavior over the spin-1 unit, which are fully supported by our many-body calculations. Additionally, magnetic-field dependent STS shows a considerably renormalized $g$-factor consistent with the multi-channel behavior, predicted by our theory.

\begin{figure}
    \centering
    \includegraphics[width=\linewidth]{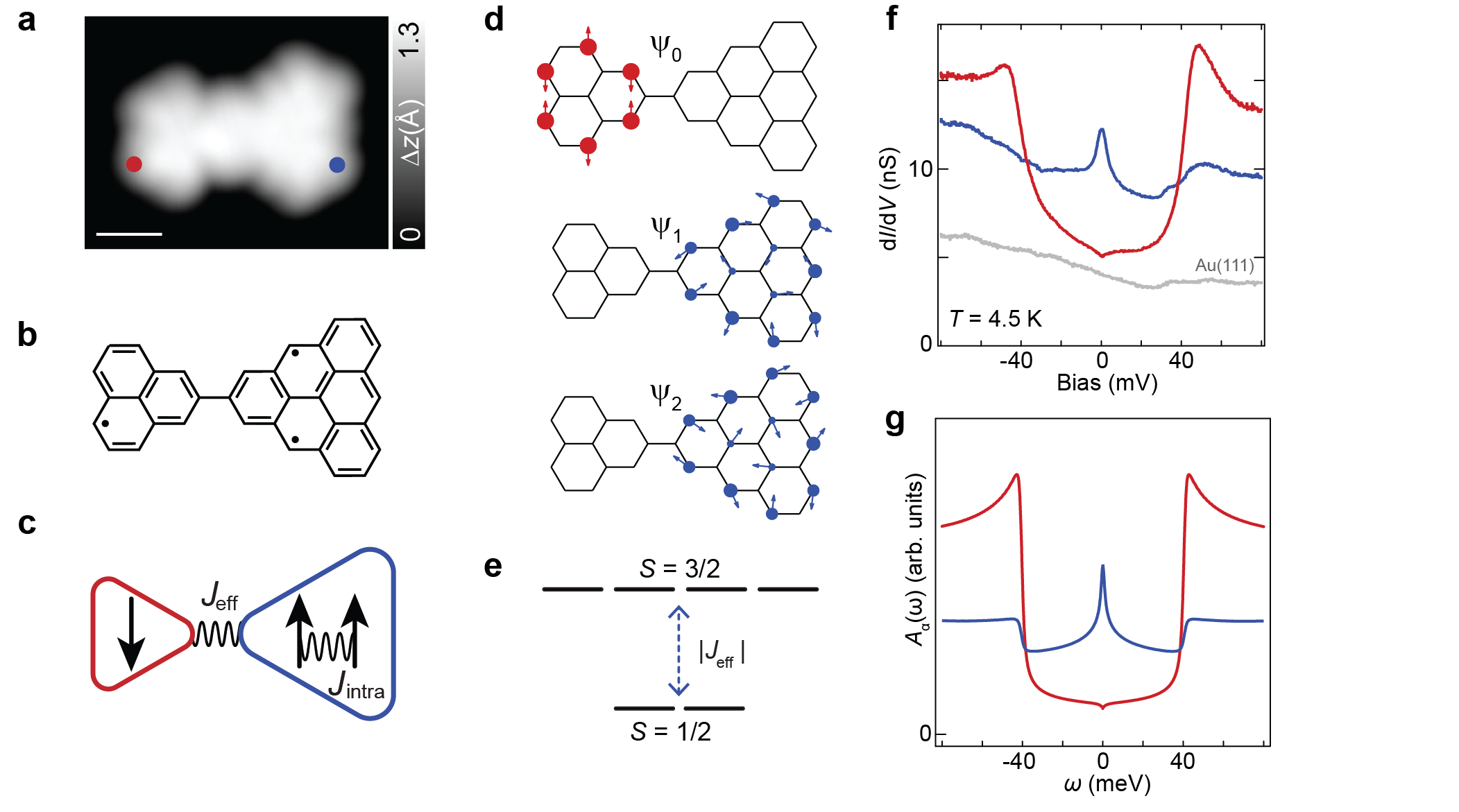}
    \caption{\textbf{Magnetic characterization of 2T-3T dimer}.
    \textbf{a} STM image of an isolated 2T-3T dimer on Au(111) ($V = -0.1$ V, $I = 100 $pA, scale bar = 0.5 nm). \textbf{b,c} Chemical structure and Heisenberg representation for the 2T-3T dimer. \textbf{d} Representation of the three zero modes, where the circle size indicates the absolute value of the wave function, and the arrows encode the complex phase. \textbf{e} Magnetic spectrum derived from the analytical solution of the Heisenberg dimer model for an antiferromagnetic coupling between the 2T and 3T units. \textbf{f} Low-bias STS spectra acquired with a carbon monoxide functionalized tip, on two distinct locations on the molecule, indicated in Fig.~\ref{fig:intro}a with colored filled circles. The reference spectrum, taken with the same tip on bare Au(111), is offset by -6\,nS (grey line). 
    Open feedback parameters \didv spectra: $V=-80$ mV, $I=850$ pA; Lock-in modulation $V_{\rm m}=1 $ mV; $T_\text{sample}=4.5$ K. \textbf{g} Orbital-resolved spectral function calculated with the one-crossing approximation at $T=4.6$ K for particle-hole symmetry and the interactions given by (\ref{eq:interactions}). The red and blue colors denote the 2T and 3T units, respectively.}
    \label{fig:intro}
\end{figure}

\section*{Results}

The on-surface synthesis of atomically precise nanographenes offers a unique platform for constructing designer spin systems~\cite{deOteyzaFrederiksen2022}, where the topology of the $\pi$-system defines the total spin ground state (GS), while the spatial symmetry governs the hybridization of these spins with the metallic substrate~\cite{Calvo-Fernandez:PRB:2024}. Here, by combining low-temperature scanning probe microscopy experiments with theoretical modeling, we demonstrate that the asymmetric spin-(1/2,1) nanographene dimer shown in Fig. \ref{fig:intro}b simultaneously realizes ferromagnetic Kondo and overscreened Kondo effects when adsorbed on the Au(111) surface. The triradical molecule under investigation arises from the covalent coupling of zigzag-edged triangular nanographenes, termed $[n]$triangulenes ($n$T), where $n$ denotes the number of benzene rings along one edge.
In this notation, the molecule shown in Fig.~\ref{fig:intro}b is a 2T-3T dimer. 
Triangulenes are (multi)radicals and thus magnetic~\cite{Turco:JACSAu:2023, SuTelychkoSongLu2020}. Their ground state (GS) spin $S$
stemming from unpaired electron(s) in the $\pi$-orbitals of the carbon atoms, 
can be inferred from the Ovchinnikov-Lieb rules~\cite{Ovchinnikov:TCA:1978,Lieb:PRL:1989}, yielding $S=(n-1)/2$. Upon covalent coupling, an effective antiferromagnetic Heisenberg exchange interaction between the spin-1/2 and spin-1 units emerges (see Fig.\ref{fig:intro}c ), stabilizing a spin-1/2 GS for the dimer---an assignment confirmed by our model calculations below. 

While a comprehensive electronic and magnetic characterization of the 2T-3T dimer is detailed in Ref.~\citenum{turco_arxiv}, our focus here is on the low-energy spectroscopic features to elucidate Kondo correlations.
To this end, we employ low-bias scanning tunneling spectroscopy (STS) on a single molecule, displayed in the scanning tunneling microscopy (STM) image in Fig. \ref{fig:intro}a. The STS spectra in Fig. \ref{fig:intro}f were taken at two different locations over the molecule marked by filled circles
in Fig. \ref{fig:intro}a, one over the 2T unit (red) and one over the 3T unit (blue). 
The \didv spectrum acquired on the 2T unit (red line) shows two conductance steps symmetric
around zero bias at $V\sim\pm40$ meV, typically indicative of inelastic spin excitations. 
Additionally, there is a weak but clearly discernible dip feature at zero bias.
Such a dip is a hallmark of the ferromagnetic Kondo effect~\cite{Koller:PRB:2005,Baruselli:PRB:2013}.
A conductance dip of a few meV could in principle also be ascribed to vibrational excitation of the carbon monoxide at the tip apex. However, the zero-bias dip can also be observed with a metal tip, as shown later in the manuscript (Fig~\ref{fig:B-field IETS}c).
Further evidence that the dip originates from the spin-1/2 GS of the molecule comes
from the fact that it vanishes if one of the spins on the 3T unit is quenched by
hydrogenation, thus leading to a total spin $S=0$ GS of the dimer (see 2T-H3T dimer in Fig.~S7 in the Supplementary Information (SI)).
On the other hand, over the 3T unit, the \didv shows a Kondo-like peak at zero-bias, accompanied by two broadened step features at the same energy, compared to those observed on the 2T unit, but with lower intensity. We now turn to theoretical modeling to interpret the observed spectroscopic signatures.

\subsection*{Microscopic modeling and effective Kondo description}

To rationalize the low-energy spectroscopic features observed in the 2T–3T nanographene dimer, we model the system using a Hubbard Hamiltonian for the $\pi$-orbitals of the carbon atoms and project onto the zero-energy spin-degenerate eigenmodes (ZMs) of the individual triangulene units~\cite{Ortiz:NL:2019,Jacob:PRB:2021,Jacob:PRB:2022,Krane:NL:2023}. This projection captures the essential low-energy physics, including spin excitations and Kondo correlations, while drastically reducing the complexity of the problem, as we show in the SI.

For the 2T-3T dimer, there are three ZMs, one ($\psi_0$) on the 2T-unit, and two ($\psi_+,\psi_-$)
on the $3T$-unit, as shown in Fig.~\ref{fig:intro}d.
Using the $C_3$--symmetric ZMs as the basis for the 3T unit~\cite{Ortiz:2DM:2022}, the interaction part of the projected Hamiltonian simplifies (see Methods) and the resulting three-ZM Hamiltonian with effective 2T–3T exchange reads:
\begin{equation}
    \label{eq:H_imp}
    \mathcal{H}_{\rm imp} = \sum_{\alpha=0,+,-} \left( \varepsilon_\alpha \, \hat{n}_\alpha + \mathcal{U}_\alpha \,  \hat{n}_{\alpha\up} \hat{n}_{\alpha\dn} \right) + \mathcal{U}^\prime \, \hat{n}_+ \hat{n}_- - J_{\rm H} \, \hat{\mathbf{S}}_+ \cdot \hat{\mathbf{S}}_- 
    + \frac{2}{3}J_{\rm eff} \, \hat{\mathbf{S}}_0 \cdot ( \hat{\mathbf{S}}_+ + \hat{\mathbf{S}}_- )
\end{equation}
where the subscripts $\alpha=0,+,-$ refer to the three ZMs ($\psi_0$, $\psi_+$, $\psi_-$), 
defined above and shown in Fig.~\ref{fig:intro}d, $\hat{n}_{\alpha\sigma}$ is the occupation number operator for ZM $\alpha$ and spin 
$\sigma=\up,\dn$, $\hat{n}_\alpha=\hat{n}_{\alpha\up}+\hat{n}{_\alpha\dn}$, and
$\hat{\bm{S}}_\alpha$ is the total spin operator for ZM $\alpha$. $\varepsilon_\alpha$ are the energy levels of the ZMs (relative to the Fermi level of the substrate),
$\mathcal{U}_\alpha$ is the intra-orbital Coulomb repulsion for ZM $\psi_\alpha$, and $\mathcal{U}^\prime$ and 
${J}_{\rm H}$ are, respectively, the inter-orbital Coulomb repulsion and the direct exchange 
(or Hund's rule coupling)  between the two ZMs on the 3T unit.
${J_{\rm eff}}$ is an effective exchange coupling between the 2T and 3T units, capturing
both kinetic and Coulomb-driven superexchange~\cite{Jacob:PRB:2022}. 
 Numerical values for the interactions in (\ref{eq:H_imp}) are
given in (\ref{eq:interactions}) in the Methods section.

Diagonalization of $ \mathcal{H}_{\rm imp}$ yields a doubly-degenerate
GS manifold with total spin $S=1/2$, consistent with Ovchinnikov-Lieb rules:
\begin{equation}
  \label{eq:gs_wf}
  |\chi_\sigma\rangle = \left\{
  \begin{array}{cc}  
    \tfrac{1}{\sqrt3} \left( \ket{\up}_0 \, \ket{\Rt}_{\pm} - \sqrt{\tfrac{2}{3}}\,\ket{\dn}_0 \, \ket{\Up}_{\pm} \right) & \mbox{ for } \sigma=\up
    \vspace{2ex} \\
    \tfrac{1}{\sqrt3} \left( \ket{\dn}_0 \, \ket{\Rt}_{\pm} - \sqrt{\tfrac{2}{3}} \, \ket{\up}_0 \, \ket{\Dn}_{\pm} \right) & \mbox{ for } \sigma=\dn 
  \end{array}
  \right.
\end{equation}
where $\ket{\Up}_{\pm},\ket{\Rt}_{\pm},\ket{\Dn}_{\pm}$ denote the spin-triplet states ($S_{\pm}=1$) with
$S^z_{\pm}=+1,0,-1$ formed by the two ZMs $\psi_+,\psi_-$ on the 3T-unit. The GS wavefunction
$\ket{\chi_\sigma}$ thus represents an entangled state between the spin-1/2 of the 2T unit and the
spin-1 of the 3T unit.

Coupling of the ZMs to the conduction electrons in the substrate yields a three-orbital
Anderson impurity model with the three ZMs as impurity levels~\cite{Jacob:PRB:2021,Krane:NL:2023,Calvo-Fernandez:PRB:2024}. 
The coupling to the substrate gives rise to a broadening of the impurity levels
$\Gamma_\alpha=\pi \,|V_\alpha|^2 \, \rho_c$ where $\rho_c$ is 
the conduction electron density of states and $V_\alpha$ is the coupling between a ZM $\psi_\alpha$ and the conduction electron states 
around the Fermi level (assumed to be constant). 
Our density functional theory calculations yield a coupling strength $\Gamma_0\sim55\,\meV$
for the 2T unit and equal couplings $\Gamma_+=\Gamma_-\sim35\,\meV$ for 
the two ZMs $\psi_+$ and $\psi_-$ on the 3T unit. Coupling between the 3T ZMs  
is negligible, indicating that each ZM interacts with a distinct 
orthogonal conduction channel.

We solve the Anderson impurity model within the one-crossing approximation~\cite{Haule:PRB:2001},
which consists in a diagrammatic expansion of the Greens function for the many-body eigenstates of the isolated impurity in the coupling to the conduction electrons in the substrate, and yields the spectral function for the coupled impurity (see Methods for details).
The spectral function is directly related to the $dI/dV$ measured in STS experiments~\cite{Jacob:PRB:2021}.
Fig.~\ref{fig:intro}g shows the calculated spectral functions $A_\alpha(\omega)$
of the ZM $\alpha=0$ on the 2T unit (red line)
and the two degenerate ZMs $\alpha=+,-$ on the 3T unit (blue line). 

The 2T spectrum exhibits a zero-bias dip accompanied by step features at $\omega\sim\pm40$ meV, corresponding to an inelastic spin excitation to $S=3/2$ (Fig. \ref{fig:intro}e). The 3T spectrum displays a zero-bias Kondo resonance with symmetric steps at the same energies, with the Kondo intensity markedly reduced relative to the 2T inelastic features. The OCA calculations indicate that this suppression arises from non-Fermi-liquid behavior rather than thermal broadening, consistent with an overscreened Kondo effect. Overall, the calculated spectral functions of the interacting ZMs coupled to the substrate conduction electrons reproduce the experimental spectroscopic phenomenology.
We note that the observed bias asymmetry especially in the 2T spectra may be attributed to
a slight detuning from particle-hole symmetry, see Sec.~III and Fig.~S3 of the SI for details.

\begin{figure}[ht!]
    \centering
    \includegraphics[width=\linewidth]{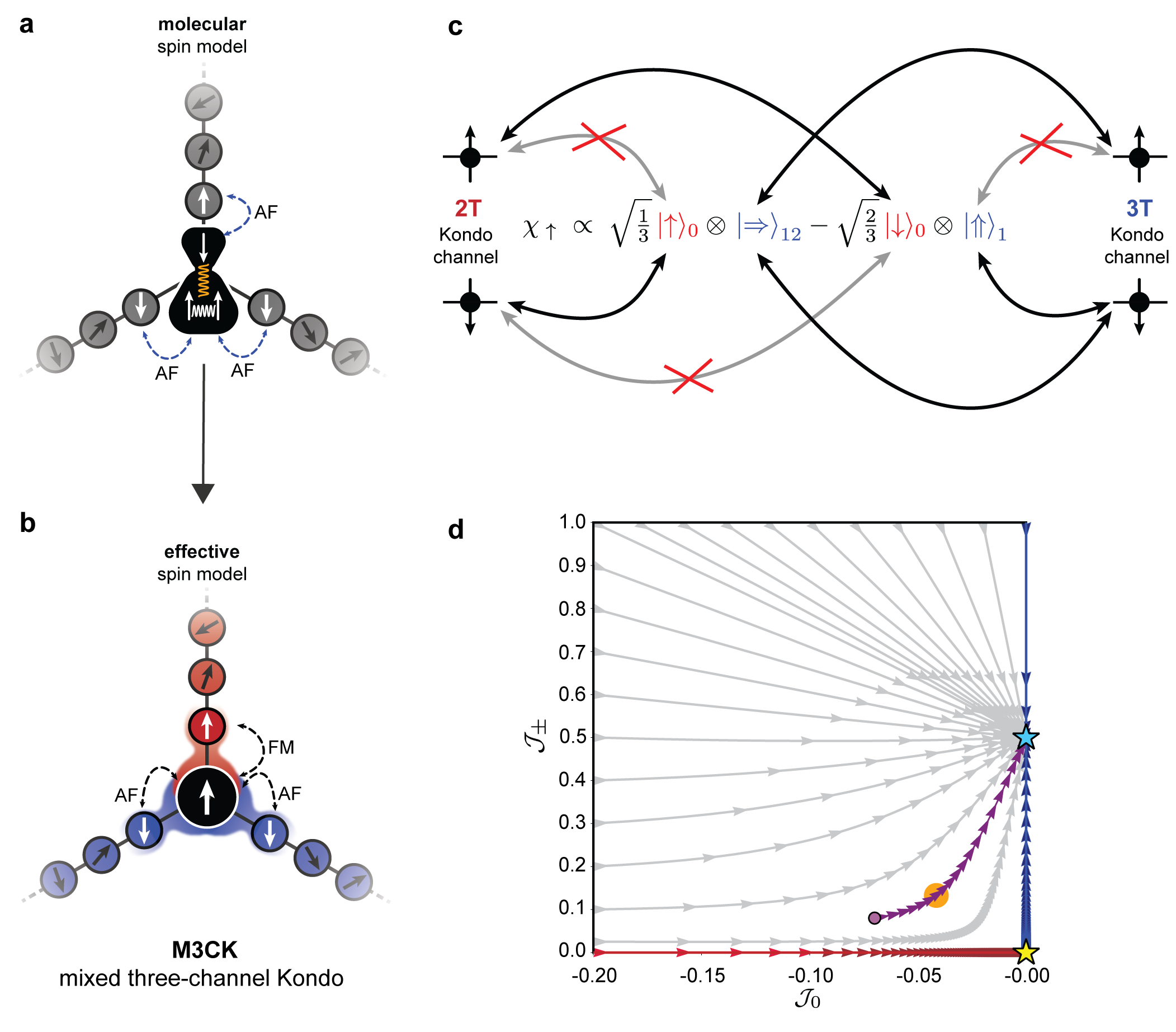}
    \caption{
    \textbf{Theoretical description of the Kondo effect in the 2T-3T dimer}.
    \textbf{a,b} Illustrations of the two models employed here for the description of 2T-3T dimer 
    coupled to conduction electrons:
    \textbf{a} Three-orbital Anderson impurity model of 2T-3T dimer
    coupled to three independent conduction electron baths, Eqs.~(\ref{eq:H_imp},\ref{eq:H_AIM}); \textbf{b} effective 
    Kondo model of spin-1/2 with three screening channels, Eqs.~(\ref{eq:H_K},\ref{eq:SW}).
     \textbf{c} Illustrations of the exchange processes between the dimer and substrate conduction electrons mediated by the 2T and 3T units, illustrating ferromagnetic and antiferromagnetic coupling mechanisms.
    \textbf{d} Third-order renormalization-group scaling trajectories of the exchange couplings $\CJ_0$ and $\CJ_\pm$ for the M3CK model, Eq.~(\ref{eq:scaling}).
    The red trajectory corresponds to a  ferromagnetic one-channel Kondo model, 
    and the blue trajectory to an antiferromagnetic two-channel Kondo model. 
    The purple trajectory denotes the physical scaling flow relevant to the 2T--3T dimer, 
    determined by the initial conditions in Eq. (\ref{eq:initial-condition}) (purple circle).
    The yellow star marks the weak coupling fixed point of the ferromagnetic Kondo effect, while 
    light blue star indicates the M3CK fixed point, characterized by the coexistence of 
    ferromagnetic Kondo effect in the 2T channel and an overscreened Kondo effect in the 3T channels. The orange filled circle denotes the cutoff corresponding to the experimental situation ($T_{\rm sample}\sim4.5$K and $V_{\rm m}\sim1$mV).
    }
    \label{fig:theory}
\end{figure}

In order to further analyze the low-energy phenomena, we map the Anderson impurity model to an effective Kondo exchange model for the molecular spin-1/2 by means of a Schrieffer-Wolff transformation to second order in the coupling to the conduction electrons. Since the three ZMs of the molecule couple to mutually
orthogonal sets of conduction electrons in the substrate, the Schrieffer-Wolff transformation yields a three-channel Kondo model, one conduction electron channel for each of the three ZMs (see Methods for details):
\begin{equation}
    \label{eq:H_K} 
    \HK = \mathcal{H}_{\rm c} + 2 \, \sum_{\alpha} \JK{\alpha} \, \bm{S}  \cdot \bm{s}_{\alpha}  
\end{equation}
The first term in (\ref{eq:H_K}) is the kinetic energy of the three conduction electron channels.
The second term is the Kondo exchange interaction between the molecular spin-1/2 
and the three conduction electron channels, where $\bm{S}$ is the molecular spin operator, and 
$\bm{s}_\alpha$ are the spin operators for each conduction electron channel $\alpha$.
We find that the signs of the effective exchange couplings $J_\alpha$, obtained by Schrieffer-Wolff transformation from the 
Anderson impurity model, depend on whether the coupling occurs via the 2T ($\alpha=0$) or the 3T unit ($\alpha=\pm$). 
More precisely, the exchange coupling via the 2T unit is ferromagnetic ($J_0<0$), 
while the exchange coupling via the 3T is antiferromagnetic ($J_\pm>0$):
\begin{equation}
    \label{eq:SW}
    \JK{0} = -\tfrac{4}{3}\, \bar{V}_0^2/\delta{E}_0 < 0
    \hspace{1ex}\mbox{ and }\hspace{1ex}
    \JK{\pm} = \tfrac{8}{3} \, \bar{V}_\pm^2 / \delta{E}_\pm > 0
\end{equation}
where $\bar{V}_\alpha$ is the average coupling of ZM $\psi_\alpha$ to the conduction electrons,
and $\delta{E}_\alpha$ are the corresponding charging energies for adding or removing one electron 
to a ZM on either the 2T ($\delta{E}_0$) or the 3T unit ($\delta{E}_\pm$), given by
Eq.~(\ref{eq:charging-energies}) in Methods.

The opposite signs of the exchange couplings follow directly from the Clebsch--Gordan structure of the ground-state wavefunction in Eq.~(\ref{eq:gs_wf}).
As illustrated in Fig.~\ref{fig:theory}c, exchange via the 2T unit is dominated by the larger-amplitude second component ($\sim \sqrt{2/3}$), which mediates ferromagnetic coupling, while the smaller first component ($\sim \sqrt{1/3}$) contributes an antiferromagnetic term. The net interaction is therefore ferromagnetic ($J_0 < 0$; Fig.~\ref{fig:theory}c). For the 3T unit, the hierarchy is reversed: ferromagnetic exchange arises only from the smaller component, whereas antiferromagnetic exchange is allowed through both components. Consequently, the overall coupling is antiferromagnetic ($J_{\pm} > 0$).

Coupling of the conduction electrons through the 2T unit alone realizes a ferromagnetic single-channel Kondo model, giving rise to the ferromagnetic Kondo effect~\cite{Koller:PRB:2005,Baruselli:PRL:2013}. In contrast, coupling through the 3T unit alone realizes a two-channel Kondo model, leading to the overscreened Kondo effect~\cite{Nozieres:JP:1980}. The experimentally relevant case of the 2T--3T dimer on the Au(111) surface, however, involves simultaneous coupling via both units, thus realizing a three-channel Kondo model with mixed ferromagnetic and antiferromagnetic couplings (M3CK). 

To elucidate the resulting competition, we employ Anderson's poor man's scaling~\cite{Anderson:JPCSSP:1970} to determine the low-energy flow of the M3CK model defined in Eqs.~(\ref{eq:H_K}) and (\ref{eq:SW}).
In this approach, high-energy conduction electrons are successively integrated out,
whereby the coupling between low- and high-energy degrees of freedom is treated perturbatively to finite order, leading to a renormalization of the exchange couplings \( J_\alpha \). The poor-man scaling procedure yields a set of coupled differential equations, called scaling equations, that describe the renormalization-group flow of \( J_\alpha \) as the conduction electron bandwidth cutoff \( \Lambda \) is progressively reduced. Physically, this reduction corresponds to lowering the temperature or bias voltage in STM experiments. To third order, the resulting scaling equations for the M3CK model read (see Methods for details):
\begin{equation}
\label{eq:scaling}
    \frac{d \CJ_\alpha}{d\ln\Lambda} = -2 \, \CJ_\alpha^2 + 2\sum_{\alpha'} \CJ_{\alpha^\prime}^2 \, \CJ_\alpha
\end{equation}
Here, we have introduced the dimensionless exchange couplings \( \CJ_\alpha \equiv \rho_c J_\alpha \), defined by multiplication with the conduction electron density of states \( \rho_c \). The first term on the right-hand side of Eq.~(\ref{eq:scaling}) represents the second-order contribution, which accounts only for the renormalization within each channel \( \alpha \). The coupling between different channels arises only at third order, captured by the second term on the right-hand side of Eq.~(\ref{eq:scaling}).

Fig.~\ref{fig:theory}d shows representative solutions $\CJ_0(\Lambda)$ and $\CJ_\pm(\Lambda)$ of Eq.~(\ref{eq:scaling}), 
called scaling trajectories for different initial conditions. 
Arrows indicate the renormalization flow as the cutoff $\Lambda$ is reduced.
For pure ferromagnetic coupling ($\CJ_0<0$, $\CJ_\pm=0$), the trajectory (red) flows to the so-called 
weak-coupling fixed point (yellow star), $\CJ_0\rightarrow0$, associated with the ferromagnetic Kondo effect. In this limit, the molecular spin ultimately decouples from the conduction electrons and becomes asymptotically free~\cite{Koller:PRB:2005}.
For purely antiferromagnetic coupling ($\CJ_0=0$, $\CJ_\pm>0$), the trajectories (blue) 
flow to the so-called intermediate-coupling fixed point (light blue star), associated with the (two-channel) overscreened 
Kondo effect, in which the molecular spin is overscreened by two competing screening channels.
For combined ferromagnetic and antiferromagnetic coupling ($\CJ_0<0$, $\CJ_\pm>0$), all trajectories (gray) flow 
towards weak-coupling for the 2T-derived channel and intermediate-coupling for the 3T-derived channel,
$(\CJ_0,\CJ_\pm)\to(0,\tfrac{1}{2})$, which we refer to as the M3CK fixed point $\CJ^\ast_{\rm M3CK}$. 
The experimentally relevant trajectory (purple) is given by the initial conditions of 
Eq.~(\ref{eq:initial-condition}), corresponding to the situation of the 2T--3T dimer on Au(111).
A more detailed discussion of the scaling theory results is given in Methods.

Overall, the scaling analysis demonstrates that ferromagnetic and overscreened Kondo physics coexist in the M3CK model capturing the low-energy physics of the 2T--3T dimer on the Au(111) surface.
The calculated spectral features are therefore consistent with the concurrent 
manifestation of ferromagnetic Kondo and overscreened Kondo signatures in the system.

\subsection*{Magnetic‑field evolution and renormalized g‑factor}

The multi-channel nature of the M3CK model directly affects the magnetic-field dependence of the spectroscopic features. In the conventional single-channel Kondo effect, the strong-coupling regime ($T,B<\TK$) suppresses the magnetic response due to formation of the Kondo singlet, such that the effective $g$-factor is strongly reduced below the characteristic crossover scale $B_c\sim \TK$~\cite{Costi:PRL:2000,Garst:PRB:2005}. By contrast, overscreening in a multi-channel system, induces magnetic frustration, giving rise to a finite, renormalized response~\cite{Horig:PRB:2012}. Consequently, one expects (i) the crossover scale associated with singlet breaking to vanish, $B_c\rightarrow 0$, and (ii) the effective $g$-factor to deviate from its bare value ($g\approx2$ for Au(111)~\cite{zhang_temperature_2013}) and to evolve continuously with the applied field $B$.

\begin{figure}[h!]
    \centering
    \includegraphics[width=\linewidth]{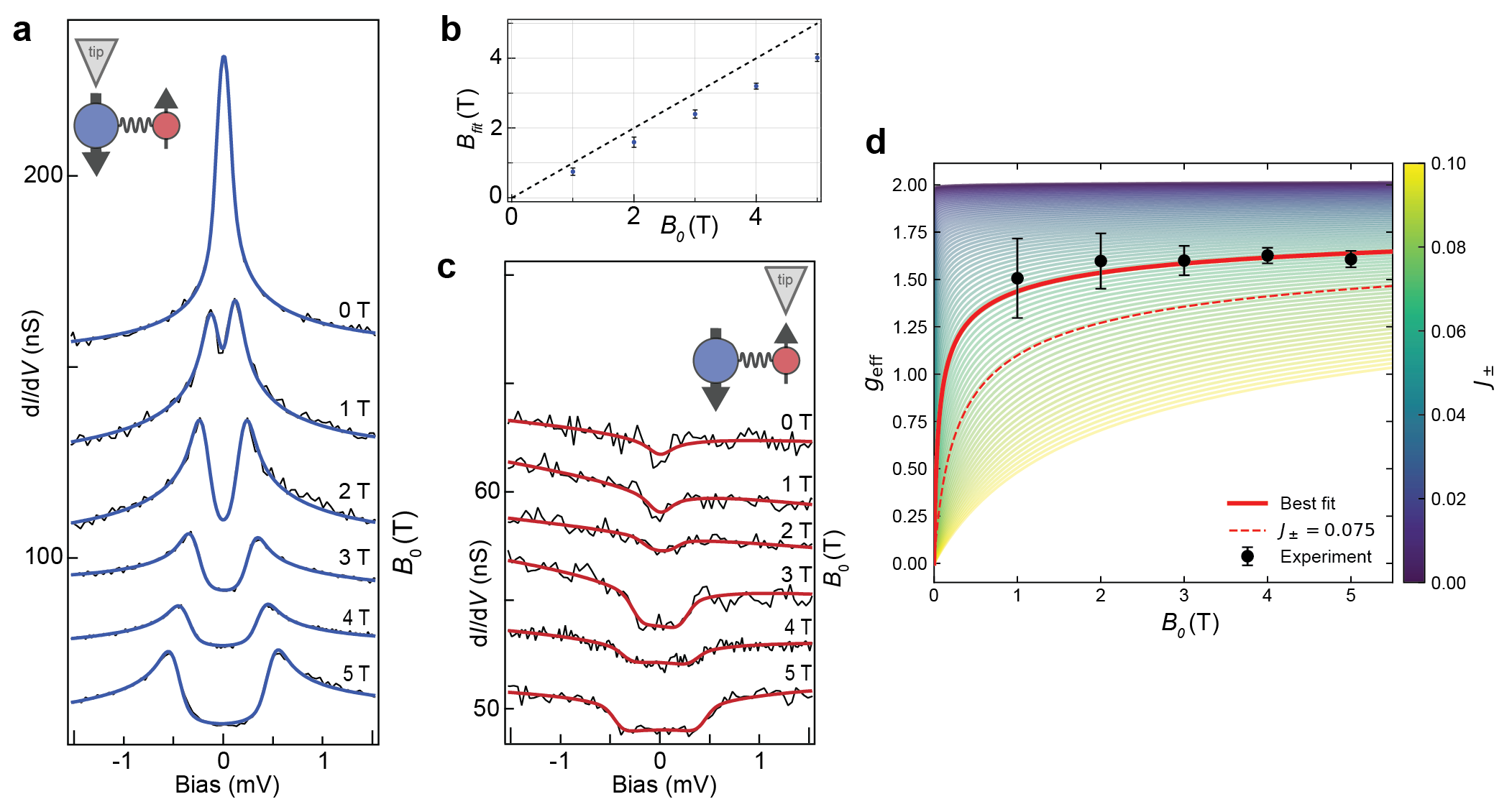}
    \caption{\textbf{Magnetic-field-dependent splitting of the overscreened Kondo peak and ferromagnetic Kondo dip.} 
\textbf{a,c} Series of \didv spectra (solid black line) acquired with a metal tip at 3\,T and 2\,T units, respectively, as a function of the applied magnetic field $B_0$. All spectra are normalized, to correct for dynamical Coulomb blockade (details in SI). Theoretical fit functions, using a third-order perturbative scattering model~\cite{Ternes:NJP:2015} with Frota lineshapes convolved with the Fermi–Dirac distribution to account for thermal broadening, are shown as solid lines: blue for 3T and red for 2T.
The spectra for $B_0=$ 0\,T, 1\,T, 2\,T, 3\,T, 4\,T are offset for clarity by 80\,nS, 50\,nS, 24\,nS, 32\,nS, 25\,nS in panel a and by 8\,nS, 7.5\,nS, 6\,nS, 2.3\,nS, 7\,nS in panel c. The raw data without Coulomb blockade corrections and offsets are shown in Fig.~S5.
\textbf{b} Effective magnetic field $B_\mathrm{fit}$ obtained from the fits to the magnetic field evolution of the overscreened Kondo peak (assuming $g=2$), plotted as a function of the applied magnetic field $B_0$. These values of $B$ are then held fixed when fitting the magnetic field dependence of the ferromagnetic Kondo dip shown in \textbf{c}. 
\textbf{d} Renormalized $g$-factor as a function of the applied magnetic field $B_0$, calculated for different values of Kondo exchange coupling $\CJ_\pm$, according to Eq.~(\ref{eq:renorm_g-factor}). The experimental values $g_{\mathrm{exp}}$, extracted from \textbf{b}, are shown as black filled circles with their error bars. The best fit ($\CJ_\pm \approx 0.62$) is shown as a solid red line, while the red dashed line indicates the $g$-factor corresponding to $\CJ_\pm \approx 0.75$, obtained via a Schrieffer--Wolff transformation. The fit was performed by $\chi^2$ minimization using the experimental uncertainties, yielding a reduced $\chi^2$ of 0.23. Error bars in panels b,d correspond to $\pm \sigma$. Open-feedback parameters: $V = -20$\,mV, $I = 1$\,nA; lock-in modulation $V_{\rm m} = 80$\,$\mu$V; $T_\text{sample} = 54$\,mK.
} 
    \label{fig:B-field IETS}
\end{figure}

To test these predictions, we measured the magnetic-field dependence of the spectroscopic signatures associated with ferromagnetic and overscreened Kondo effect.
Figures~\ref{fig:B-field IETS}a and~\ref{fig:B-field IETS}c show normalized \didv spectra (see SI for details), acquired at $T = 54$ mK as a function of the applied magnetic field $B_0$ for tip positions above 3T and 2T, respectively. 
The zero-field spectrum measured on 3T exhibits a narrow zero-bias resonance well described by a Frota function (solid blue line) with a half width at half maximum of 54 µeV. 

We fit the field-dependent spectra in Figure \ref{fig:B-field IETS}a using a third-order perturbative scattering model~\cite{ternes_probing_2017}. In contrast with previous reports on one-channel Kondo systems~\cite{zhang_temperature_2013, esat_electron_2023}, we required three key distinctions in our model to fit the data: (i) the resonances cannot be captured by a logarithmic function (see SI) and are best captured by a temperature-broadened Frota lineshape (solid blue lines in Fig.~\ref{fig:B-field IETS}a), (ii) a quantitatively accurate fit requires the critical field to vanish $B_c \rightarrow0$, and (iii) the effective magnetic field $B_\mathrm{fit}$, used as fit parameter in the model and assuming a bare $g=2$, deviates significantly from the applied field $B_0$, revealing a renormalized $g$-factor (Fig.~\ref{fig:B-field IETS}b).

The renormalization of the $g$-factor in overscreened Kondo systems has been derived analytically in Ref.~\citenum{horig_transport_2012}. In the present system, the 3T unit is overscreened, but the presence of a ferromagnetic Kondo channel must also be considered. 
Generalizing the second-order scaling equation for the $g$-factor of the one-channel Kondo model derived in Ref.~\citenum{Garst:PRB:2005} to the multi-channel case and integrating, we obtain
\begin{equation}
    g_{\rm eff}(\Lambda) = g\, \exp\left( 2 \int_{\Lambda_0}^\Lambda d\ln\Lambda^\prime \sum_\alpha\left(\CJ_\alpha(\Lambda^\prime)\right)^2 \right) .
\label{eq:renorm_g-factor}
\end{equation}
Here $\Lambda_0$ is set by the spin excitation energy ($\sim40\,\meV$), while the running cutoff $\Lambda$ is determined experimentally by the magnetic field ($B=1\,\mathrm{T}\sim58\,\mu\mathrm{eV} \gg kT \sim 5\,\mu\mathrm{eV}$). The initial value $g_{\rm eff}(\Lambda_0)=g_i$ corresponds to the Knight-shift corrected bare $g$-factor, $g_i = g - \sum_\alpha \CJ_\alpha$, with $g=2$~\cite{Garst:PRB:2005}.
Numerical integration of Eq.~\ref{eq:renorm_g-factor} with the scaling trajectories
$\CJ_\alpha(\Lambda)$ extracted from the model yields $g_{\rm eff}(B_0)$. 
The result is highly sensitive to the antiferromagnetic couplings $\CJ_\pm$ of the two 3T channels, while its dependence on the ferromagnetic coupling $\CJ_0$ is strongly suppressed due to approximate cancellation between first-order (Knight-shift) and higher-order contributions (see SI).
Figure~\ref{fig:B-field IETS}d compares $g_{\rm eff}(B_0)$ for varying $\CJ_\pm$ at fixed $\CJ_0=-0.07$ with the experimentally extracted $g$-factor. Good agreement is obtained for $\CJ_\pm \sim 0.62$, close to the value $\CJ_\pm \sim 0.75$ from the Schrieffer--Wolff transformation of the three-orbital Anderson model. The remaining deviation is consistent with the exponential sensitivity of Kondo scales to microscopic parameters. 

A single-channel Kondo (1CK) model fails to reproduce the observed field dependence (see SI). The renormalized $g$-factor therefore provides strong evidence for the multi-channel character of the 2T--3T dimer on Au(111), consistent with the M3CK description.

Having established the origin of the renormalized $g$-factor on the 3T unit, we turn to the 2T unit as an independent test of the M3CK model and a direct probe of the ferromagnetic Kondo regime. 
The zero-field spectrum exhibits a narrow zero-bias dip, consistent with the predicted $1/\ln^2({eV/k_{\rm B}T_0})$ singularity of the ferromagnetic Kondo~\cite{Koller:PRB:2005,Baruselli:PRL:2013}. 
Upon application of a magnetic field, the zero-bias anomaly is suppressed, giving rise to a symmetric pair of inelastic Zeeman steps that are quantitatively reproduced by the perturbative scattering model (solid red lines in Fig.~\ref{fig:B-field IETS}c). The resulting $\mathrm{d}I/\mathrm{d}V$ spectra show good agreement with the theoretical predictions.

Importantly, in contrast to Ref.~\citenum{Baruselli:PRB:2013}, where the spin-1/2 remains a free local moment with $g = 2$, the present system is governed by coupled ferromagnetic and overscreened Kondo correlations within the M3CK model. The field dependence of the 2T--ferromagnetic Kondo response is described using the renormalized $g$-factor extracted from the 3T sector, and the quantitative agreement establishes the multi-channel character of the dimer.

\section*{Discussion}

The 2T–3T dimer on Au(111) realizes a single spin–1/2 coupled to three independent conduction channels with opposite exchange symmetry. Two channels (from the 3T unit) couple antiferromagnetically and drive the system into an overscreened, non-Fermi-liquid regime, while the third (from the 2T unit) couples ferromagnetically and produces the hallmark singular Fermi-liquid response. These distinct signatures - suppressed zero-bias resonance and logarithmic zero-bias dip - are observed simultaneously within a single molecule and evolve characteristically under magnetic field.

This coexistence of overscreened and ferromagnetic Kondo physics establishes a mixed-channel Kondo state that cannot be realized in conventional single-impurity systems. The impurity spin remains only partially screened, yielding a robust residual spin–1/2 and directly revealing the competition between distinct many-body screening mechanisms.

More broadly, our results demonstrate that atomically precise nanographenes enable deterministic engineering of quantum impurity Hamiltonians, including the number of screening channels and the sign of the exchange interaction. This capability opens a route to controlled realization of non-Fermi-liquid states and other exotic boundary phenomena in a real-space platform. Extending this approach to coupled impurities and superconducting environments provides direct access to collective screening, quantum critical behavior~\cite{Jarrell1996,Ge2024}, and the interplay between multi-channel Kondo physics and Yu–Shiba–Rusinov states~\cite{Franke2011,PhysRevB.95.085121}.

\section*{Methods}

\subsection*{Effective three-orbital model of 2T--3T dimer}

The Hamiltonian (\ref{eq:H_imp}) is an effective three-orbital model for the 2T--3T dimer that goes beyond
a simple projection of the Hubbard model onto the three ZMs, since it also includes the effective super-exchange term 
$\frac{2}{3}J_{\rm eff} \, \hat{\mathbf{S}}_0 \cdot ( \hat{\mathbf{S}}_+ +  \hat{\mathbf{S}}_- )$ which 
accounts not only for kinetic exchange (due to hopping between ZMs on different units) 
but also for Coulomb-driven super-exchange (due to Coulomb interactions between the ZMs and
other molecular orbitals)~\cite{Jacob:PRB:2021,Jacob:PRB:2022}. 
While the latter is completely absent in the 
CAS(3,3) calculation, the former is taken into account for CAS(3,3) including third-neighbor hopping ($t_3\neq0$).
In order to reproduce the spin excitation energy of $\sim47\,\meV$ found by exact diagonalization of the Hubbard model projected onto three ZMs and two additional molecular orbitals $U=4\,\eV$, $t_1=-2.7\,\eV$ and $t_3=0.07\,t_1$ (see SI for details), we set $J_{\rm eff}=47\,\meV$.

For the single ZM $\psi_0$ of the 2T unit there is only one interaction parameter, the intra-orbital Coulomb repulsion:
\[
\mathcal{U}_0=\bra{\psi_0,\psi_0}\mathcal{H}_U\ket{\psi_0,\psi_0}=U\sum_i|\psi_0(i)|^4=U/6 \,\sim667\,\meV.
\]
This follows from the wavefuntion $\psi_0(i)=\bracket{i}{\psi_0}$ having equal contributions from exactly 6 sites of the 2T unit.

For the two ZMs of the 3T unit we choose the $C_3$-symmetric representation of the ZMs, i.e., the eigenstates $\ket{\psi_\pm}$ of the counter-clockwise rotation operator $R_{2\pi/3}$, i.e., $R_{2\pi/3} \ket{\psi_\pm} = e^{\pm i2\pi/3} \ket{\psi_\pm}$,
see Ref.~\citenum{Ortiz:2DM:2022}.
In this basis both ZMs have exactly the same weights $w_i=|\psi_\pm(i)|=|\bracket{i}{\psi_\pm}|$ 
at each site $i$ of the 3T unit, as shown by the circle sizes in Fig.~\ref{fig:intro}d, but different phases $e^{i\theta_i}$ (shown by the arrows).
For this reason, all direct repulsion terms $\bra{\psi_k,\psi_l}\mathcal{H}_U\ket{\psi_k,\psi_l}$ for the 3T ZMs $\psi_\pm$ have the same value: 
\[
\mathcal{U}_\pm=\mathcal{U}^\prime=\bra{\psi_k,\psi_l}\mathcal{H}_U\ket{\psi_k,\psi_l} = U \sum_i |\psi_k(i)|^2 |\psi_l(i)|^2 = U \sum_i |\psi_k(i)|^4 = \frac{35}{363}U \, ,
\]
where the last step follows from the ZM wave function $\psi_k(i)=w_i e^{iw_i\theta_i}$ having 6 weights $w_i=1/\sqrt{11}$, 3 weights $w_i=2/\sqrt{33}$ and 3 weights $w_i=1/\sqrt{33}$ so that 
$\sum_i |\psi_k(i)|^4=\sum_i w_i^4=6\times(\tfrac{1}{11})^2+3\times(\tfrac{4}{33})^2+3\times(\tfrac{1}{33})^2=\tfrac{35}{363}$.
Moreover, also the direct exchange or Hund's rule coupling $J_{\rm H}\equiv\bra{\psi_k,\psi_l}\mathcal{H}_U\ket{\psi_l,\psi_k}$ (with $k\neq l$) has the same value as the direct interactions, 
since 
\[
    J_{\rm H} =  2\,U \sum_i \psi_k^\ast(i) \psi^\ast_l(i) \psi_l(i) \psi_k(i) =  2\,U \sum_i |\psi_k(i)|^4 
    =  2\,\mathcal{U}_\pm 
\]
Finally, while in general there is also a so-called pair-hopping term $\mathcal{P}_{k,l}=\bra{\psi_k,\psi_k}\mathcal{H}_U\ket{\psi_l,\psi_l}$ between the ZMs to be considered~\cite{Ortiz:NL:2019}, it is straight-forward to show that in the $C_3$-symmetric representation of the ZMs, this term vanishes: The transformation from the real ZMs $\psi_1,\psi_2$ (shown in Fig.~S1b in the SI) to the complex $C_3$-symmetric ZMs $\psi_+,\psi_-$ is given by $\psi_\pm=\frac{1}{\sqrt2}(\psi_1\pm i\psi_2)$. 
Hence in the $\psi_\pm$ basis, we have 
\[
\mathcal{P}_{+,-} = \frac{1}{4} \left( 2\,\mathcal{U}_{1111} -2\, \mathcal{U}_{1122} - 2\,\mathcal{U}_{1212} - 2\,\mathcal{U}_{1212} \right) = 0,
\]
since $\mathcal{U}_{1122}=\mathcal{U}_{1212}=\mathcal{U}_{1221}=\frac{1}{3}\,\mathcal{U}_{1111}$ in the real ZM basis, where we have defined $\mathcal{U}_{klmn}=\bra{\psi_k,\psi_l}\mathcal{H}_U\ket{\psi_m,\psi_n}$.
Thus, on the 3T unit $\mathcal{U}_\pm = \mathcal{U}^\prime = J_{\rm H}/2 = \frac{35}{363}U$, while all other interactions are zero.

Assuming $U=4\,\eV$ for the Hubbard interaction, the non-zero interactions of the three-orbital model (\ref{eq:H_imp}) then are:
\begin{equation}
    \label{eq:interactions}
    \mathcal{U}_0 = \frac{U}{6}\sim667\,\meV, \hspace{2ex}
    \mathcal{U}_\pm = \mathcal{U}^\prime = \frac{J_{\rm H}}{2} = \frac{35}{363}\,U \sim 386\,\meV, \hspace{2ex} 
    J_{\rm eff}\sim47\,\meV.
\end{equation}
The ratio between the interactions on the 3T unit and the 2T unit is thus $\mathcal{U}_\pm/\mathcal{U}_0=\frac{70}{121}\sim0.58$, i.e. the interactions on the 3T unit are smaller than on the 2T unit owing to the extension of the 3T ZMs $\psi_\pm$ over a larger number of sites than the 2T ZM $\psi_0$. 

For the Schrieffer-Wolff transformation we need the charging energies of the 2T and 3T units, i.e. the energies 
for adding or removing one electron.
Assuming particle-hole symmetry, i.e., $\mathcal{E}_0=-\frac{1}{2}\mathcal{U}_0$ for the 2T unit 
and $\mathcal{E}_\pm=-\frac{1}{2}\mathcal{U}_\pm-\mathcal{U}^\prime=-\frac{3}{2}\mathcal{U}_\pm$
for the 3T unit, the charging energies are 
\begin{equation}
    \label{eq:charging-energies}
    \delta E_0=\frac{\mathcal{U}_0}{2}\sim334\,\meV \hspace{1ex}\mbox{ and } \hspace{1ex}
    \delta E_\pm=\frac{\mathcal{U}_\pm}{2}+\frac{J_{\rm H}}{4} +\frac{J_{\rm eff}}{6} = \mathcal{U}_\pm +\frac{J_{\rm eff}}{6}
    \sim394\,\meV
\end{equation}
Hence the charging energies for the 2T and 3T units are very similar, despite the interactions being considerably
smaller for the 3T unit than for the 2T unit. 
The reason is of course that adding one electron/hole requires to overcome the Coulomb repulsion
with two electrons/holes for the 3T unit rather than just with one electron/hole as for the 2T unit.

\subsection*{Three-orbital Anderson impurity model}

The effective  three-orbital Anderson impurity model for the 2T--3T dimer on the Au(111) substrate can be written as
\begin{equation}
    \label{eq:H_AIM}
    \mathcal{H}_{\rm aim} = \mathcal{H}_{\rm imp} + \mathcal{H}_{\rm sub} + \mathcal{V}_{\rm hyb}   
\end{equation}
where $\mathcal{H}_{\rm imp}$ describes the impurity shell comprising the three ZMs $\{\psi_\alpha\}_{\alpha=0,\pm}$ of the 2T--3T dimer given by (\ref{eq:H_imp}), $\mathcal{H}_{\rm sub} \equiv \sum_{\alpha,k,\sigma} \varepsilon_{\alpha,k} \, \hat{n}_{\alpha,k,\sigma}$ describes the three mutually orthogonal conduction electron channels $\alpha=0,+,-$ in the substrate, 
and $\mathcal{V}_{\rm hyb}\equiv \sum_{\alpha,k,\sigma} V_{\alpha,k} \left( c_{\alpha,k,\sigma}^\dagger \, d_{\alpha,\sigma} + d_{\alpha,\sigma}^\dagger \, c_{\alpha k,\sigma} \right)$ the hybridization between the three ZMs with the conduction electron channels, where we have assumed that each ZM $\psi_\alpha$ couples to only one of the channels, as shown by our density functional theory calculations (see SI).
Here $\varepsilon_{\alpha,k}$ is the energy dispersion for the conduction electron states of channel $\alpha$,
$\hat{n}_{\alpha,k,\sigma}$ is the number operator, and $c_{\alpha,k,\sigma}$ ($c_{\alpha,k,\sigma}^\dagger$) is the corresponding annihilation (creation) operator for the conduction electron state in bath $\alpha$ with wavevector $k$ and spin $\sigma$. 
Integrating out the conduction electron degrees of freedom, one obtains the so-called hybridization function $\Delta_{\alpha}(\omega) \equiv \sum_{k} |V_{\alpha,k}|^2/(\omega+i\eta-\varepsilon_{\alpha,k})$, where $i\eta$ shifts the poles infinitesimally to the lower complex plain.
The real part of $\Delta_{\alpha}(\omega)$ describes the renormalization and the imaginary part the broadening  of an impurity level (i.e. the ZMs) due to the coupling to its bath.
For noble metals, we may assume an approximately constant conduction electron density of states, $\rho(\omega)\approx\rho_c$, and coupling $V_{\alpha,k}\approx{V_\alpha}$ in the vicinity of the Fermi level. In this limit, known as the wide-band limit (WBL), the hybridization function becomes constant and purely imaginary, $\Delta_\alpha(\omega)=-i\Gamma_\alpha$. The single-particle broadening $\Gamma_\alpha$ of impurity level $\alpha$ is given by $\Gamma_\alpha=\pi\,|V_\alpha|^2\rho_c$.

\subsection*{One-crossing approximation}

The Anderson impurity model (\ref{eq:H_AIM}) is solved within the one-crossing approximation~\cite{Haule:PRB:2001,Haule:PRB:2010}.
The first step is a numerical diagonalization of the isolated impurity Hamiltonian 
for the three ZMs given by (\ref{eq:H_imp}),
${\cal H}_{\rm imp}\ket{m}=E_m\ket{m}$ for different fillings $N$ of the impurity shell. 
We consider the 2T-3T dimer close to half-filling, i.e. $N=3$.
The coupling to the substrate ${\cal V}_{\rm imp-sub}$ only connects eigenstates with occupations differing 
by one electron, leading to charge and spin fluctuations in the impurity shell.
Thus for the 3 ZMs of the 2T-3T dimer we consider the occupations $N=2,3,4$.
It is the fluctuations between the impurity GS manifold and excited states with one more or
one less electron that give rise to both Kondo effects and spin excitations~\cite{Jacob:PRB:2021}.

In the next step so-called pseudo-particles (PPs) $m$ corresponding to
the many-body eigenstates $\ket{m}$ are introduced. The Green's function 
of such a PP $m$ can then be written as $G_m(\omega) = 1/(\omega-\lambda-E_m-\Sigma_m(\omega))$
where $\Sigma_m(\omega)$ is the PP self-energy which describes the
renormalization (real part) and broadening (imaginary part) of the
PP energy $E_m$ due to the interaction with other PPs $m^\prime$ mediated by
the conduction electron bath.
$-\lambda$ is the chemical potential for the PPs which has to be adjusted
such that the total PP charge $Q=\sum_ma_m^\dagger a_m$ is conserved,
in order to impose the completeness of the many-body Hilbert space.

The one-crossing approximation consists in a diagrammatic expansion of the PP self-energies $\Sigma_m(\omega)$
in terms of the hybridization function $\Delta_\alpha(\omega)$ to
infinite order but summing only a subset of diagrams (only those involving
conduction electron lines crossing at most once).
This leads to a set of coupled integral equations for the PP propagators
and self-energies that have to be solved self-consistently.
Once the one-crossing approximation is converged, the electron spectral function
$A_\alpha(\omega)$ for the impurity orbitals are obtained from convolutions of PP propagators
$G_m(\omega)$.
More details on the application of the one-crossing approximation
to Nanographenes and other nanoscale quantum magnets can be found e.g. in Refs.~\citenum{Jacob:EPJB:2016,Jacob:PRB:2021}. 

\subsection*{Derivation of the M3CK model from the Anderson impurity model by Schrieffer-Wolff transformation}

In the following we derive the Kondo exchange couplings for each of the three orbitals of our effective Anderson model, as defined in Eqs.~(\ref{eq:H_imp},\ref{eq:H_AIM}), by a Schrieffer-Wolff transformation to second order in the coupling to the conduction electrons~\cite{Schrieffer:PR:1966,Hewson:book:1997}.
The starting point is the GS of the isolated 2T--3T dimer described by the Hamiltonian (\ref{eq:H_imp}) 
which is a total spin-1/2 doublet $\{\ket{\chi_\up},\ket{\chi_\dn}\}$ given by (\ref{eq:gs_wf}).
Using the field operators $d_{\alpha,\sigma}^\dagger$ ($d_{\alpha,\sigma}$) 
for creating (destroying) an electron with spin $\sigma=\up,\dn$ in ZM $\alpha=0,+,-$ we may also write
the GS doublet as
\begin{equation}
   |\chi_\sigma\rangle =  \frac{1}{\sqrt3} \left( 
   d_{0,\sigma}^\dagger 
   \, \frac{d_{+,\up}^\dagger \, d_{-,\dn}^\dagger + d_{+,\dn}^\dagger \, d_{-,\up}^\dagger}{\sqrt2}
   -\sqrt2 \, d_{0,\bar\sigma}^\dagger \, d_{+,\sigma}^\dagger \, d_{-,\sigma}^\dagger
   \right)
   \ket{0}
\end{equation}
where $\ket{0}$ denotes the vacuum state.
The three ZMs $\psi_\alpha$ of the 2T--3T dimer coupled to the conduction electrons in the substrate 
constitute a three-orbital Anderson impurity model, described by (\ref{eq:H_AIM}). 
As established above, the coupling between substrate and impurity can be decomposed into
three individual conduction electron channels $\alpha=0,+,-$, each coupling to exactly one of the 
three ZMs $\psi_\alpha$: $\mathcal{V}_{\rm hyb}=\sum_\alpha\mathcal{V}_\alpha$. 
The coupling for the individual channels $\mathcal{V}_\alpha$ can be written in terms of an 
average coupling $\bar{V}_\alpha=\sqrt{\frac{1}{N_s}\sum_k |V_{\alpha,k}|^2}$ as
\begin{equation}
\label{eq:coupling}
  \hat{\mathcal{V}}_\alpha = \sum_{k,\sigma} V_{\alpha,k} \left( c_{\alpha,k,\sigma}^\dagger \, d_{\alpha,\sigma} 
  + d_{\alpha,\sigma}^\dagger \, c_{\alpha, k,\sigma} \right) 
  = \bar{V}_{\alpha} \sum_{\sigma} \left( \bar {c}_{\alpha,\sigma}^\dagger \, d_{\alpha,\sigma} 
  + d_{\alpha,\sigma}^\dagger \, \bar{c}_{\alpha,\sigma} \right)
\end{equation}
where $c_{\alpha,k,\sigma}^\dagger$ ($c_{\alpha,k,\sigma}$) creates (destroys) a conduction electron state 
$\phi_{\alpha,k}$ with spin $\sigma\in\{\up,\dn\}$ and wave vector $k$ in bath $\alpha$.
In the last step we have introduced $\bar{c}^\dagger_{\alpha,\sigma}$ ($\bar{c}_{\alpha,\sigma}$)
which creates (destroys) a conduction electron in the state $\ket{\bar{\phi}_{\alpha,\sigma}}\equiv \frac{1}{\bar{V}_\alpha}\sum_k V_{\alpha,k}\,\ket{\phi_{\alpha,k,\sigma}}$
localized around the corresponding ZM $\ket{\psi_\alpha}$, where $\bar{V}_\alpha$ is such that the localized conduction 
electron state $\ket{\bar{\phi}_{\alpha,\sigma}}$ is normalized, with $N_s$ the number of conduction electron states.

Coupling of the molecular spin-1/2 to the conduction electrons in the substrate
occurs via the three zero-modes of the 2T--3T dimer and thus yields three screening 
channels in the corresponding Kondo model with different Kondo couplings $\JK{\alpha}$ 
for each channel. 
We therefore consider the three-channel Kondo model given by Eq.~(\ref{eq:H_K})
in the main text. In terms of conduction electron operators the Hamiltonian can be written as
\begin{equation}
    \HK = \sum_{\alpha,k,\sigma} \varepsilon_{\alpha,k}\,n_{\alpha,k,\sigma} + 
    \sum_{\alpha} J_\alpha \left(\;
      S^+ s^-_{\alpha} + S^- s^+_{\alpha} + 2\,  S^z s^z_{\alpha} \; \right)
\end{equation}
The first term is the conduction band Hamiltonian $\mathcal{H}_{\rm c}$ comprising the three conduction electron channels, where $n_{\alpha,k,\sigma}=c^\dagger_{\alpha,k,\sigma}\,c_{\alpha,k,\sigma}$ counts the conduction electrons in channel $\alpha$ with wave vector $k$ and spin $\sigma$, and $c^\dagger_{\alpha,k,\sigma}$, $c_{\alpha,k,\sigma}$ are the corresponding creation and annihilation operators.
The second term is the Kondo exchange interaction $\mathcal{V}_{\rm K}$ between the molecular spin-1/2 and the conduction electrons in the three channels, where 
$S^+$, $S^-$ and $S^z$ are the raising, lowering and z-component operators of the molecular spin-1/2, and
\begin{equation}
s^+_{\alpha}\equiv \bar{c}^\dagger_{\alpha,\up}\,\bar{c}_{\alpha,\dn}, \hspace{1ex}
s^-_{\alpha}\equiv \bar{c}^\dagger_{\alpha,\dn}\,\bar{c}_{\alpha,\up},\hspace{1ex}\mbox{ and } \hspace{1ex}
s^z_{\alpha}\equiv \tfrac{1}{2}(\bar{c}^\dagger_{\alpha,\up}\,\bar{c}_{\alpha,\up}-\bar{c}^\dagger_{\alpha,\dn}\,\bar{c}_{\alpha,\dn})
\end{equation}
are the corresponding operators for the spins of the three conduction electron channels.

The Schrieffer-Wolff transformation from the three-orbital Anderson model to the three-channel Kondo model consists in 
treating the coupling to the conduction electrons (\ref{eq:coupling}) as a perturbation to second 
order and projecting onto the low-energy (LE) subspace formed by the doubly degenerate  GS $\{\chi_\sigma\}$ 
of the unperturbed molecule:
\begin{equation}
    \VK = \mathcal{P} \, \left[ \sum_\alpha \mathcal{V}_\alpha^\dagger \, \frac{1}{\Egs-\mathcal{H}^{(0)}} \, \mathcal{V}_\alpha \right] \, \mathcal{P}
\end{equation}
with $\mathcal{H}^{(0)}$ the Hamiltonian of the uncoupled system, and 
$\mathcal{P}$ the projector onto the LE subspace.

The exchange couplings $J_\alpha$ for each channel can be obtained from the energy difference  
between the triplet states $\ket{\Psi_{1,m}^\alpha}$ (where $m=-1,0,+1$ is the $z$-projection of the total spin $S$) 
and the singlet state $\ket{\Psi^\alpha_{0,0}}$ formed between the molecular GS doublet $\ket{\chi_\sigma}$ 
and a conduction electron in channel $\alpha$ given by $\ket{\bar\phi_{\alpha,\sigma}}$. 
Taking the $m=0$ triplet state yields:

\begin{align}
    J_\alpha &= E_{\Psi^\alpha_{1,0}} - E_{\Psi^\alpha_{0,0}} 
    = \bra{\Psi^\alpha_{1,0}}\VK\ket{\Psi^\alpha_{1,0}} - \bra{\Psi^\alpha_{0,0}}\VK\ket{\Psi^\alpha_{0,0}} 
    \nonumber\\
    &= -\bra{\chi_\up,\bar\phi_{\alpha,\dn}} \VK \ket{\chi_\dn,\bar\phi_{\alpha,\up}}
        -\bra{\chi_\dn,\bar\phi_{\alpha,\up}} \VK \ket{\chi_\up,\bar\phi_{\alpha,\dn}}
        \nonumber\\
    &= -2\bra{\chi_\up} \bar{c}_{\alpha,\dn} \, \mathcal{V}_\alpha^\dagger \, \frac{1}{\Egs-\mathcal{H}^{(0)}} \, \mathcal{V}_\alpha \,\bar{c}_{\alpha,\up}^\dagger\ket{\chi_\dn}
\end{align}
where we have used the representation of the $m=0$ triplet state as $\ket{\Psi^\alpha_{1,0}}=\tfrac{1}{\sqrt2}\left(\ket{\chi_\up}\otimes\ket{\bar\phi_{\alpha,\dn}} - \ket{\chi_\dn}\otimes\ket{\bar\phi_{\alpha,\up}}\right)$, and 
of the singlet state as
$\ket{\Psi^\alpha_{0,0}}=\tfrac{1}{\sqrt2}\left(\ket{\chi_\up}\otimes\ket{\bar\phi_{\alpha,\dn}} +\ket{\chi_\dn}\otimes\ket{\bar\phi_{\alpha,\up}} \right)$. It is worth noting that here the phases between the 
two components of the $m=0$ triplet and the singlet state are opposite w.r.t. to the usual representation, where
the spin triplet has the $(+)$-sign while the spin singlet has the $(-)$-sign, owing to the composite nature of the 
molecular spin-1/2. In the last step we have used that $\bra{\chi_\up,\bar\phi_{\alpha,\dn}} \VK \ket{\chi_\dn,\bar\phi_{\alpha,\up}}
=\bra{\chi_\dn,\bar\phi_{\alpha,\up}} \VK \ket{\chi_\up,\bar\phi_{\alpha,\dn}}$ owing to the hermeticity of $\VK$.

Thus, in order to compute $J_\alpha$ we need to compute the action of $\mathcal{V}_\alpha$ on 
$\bar{c}_{\alpha,\up}^\dagger\ket{\chi_\dn}$ and of $\mathcal{V}_\alpha^\dagger=\mathcal{V}_\alpha$ 
on $\bra{\chi_\up}\,\bar{c}_{\alpha,\dn}$. Using commutation relations, we obtain:
\begin{align}
    \mathcal{V}_\alpha \, \bar{c}_{\alpha,\up}^\dagger &= 
    \bar{V}_\alpha \, d_{\alpha,\up}^\dagger + 
    \bar{V}_\alpha\,d_{\alpha,\dn}\,\bar{c}_{\alpha,\up}^\dagger\,\bar{c}_{\alpha,\dn}^\dagger
    \\
    \mbox{ and } \; \bar{c}_{\alpha,\dn} \, \mathcal{V}_\alpha &=
    \bar{V}_\alpha \, d_{\alpha,\dn} -
    \bar{V}_\alpha\,\bar{c}_{\alpha,\dn}\,\bar{c}_{\alpha,\up}\,d_{\alpha,\up}^\dagger
\end{align}
Hence the Kondo exchange couplings $J_\alpha$ for our three-channel Kondo model are given by:
\begin{align}
    J_\alpha &= -2 \,\bar{V}_\alpha^2 \left( \bra{\chi_\up}  d_{\alpha\dn} \, \frac{1}{\Egs-\mathcal{H}^{(0)}} \, d_{\alpha\up}^\dagger \ket{\chi_\dn} - \bra{\chi_\up}  d^\dagger_{\alpha\up} \, \frac{1}{\Egs-\mathcal{H}^{(0)}} \, d_{\alpha\dn}  \ket{\chi_\dn} \right)
\end{align}
In order to proceed, we have now to specialize to the conduction channel $\alpha$ for which we want to compute
the exchange coupling.

\subsubsection*{Kondo exchange due to coupling via the 2T unit}

For the exchange coupling $J_0$ of the 2T-derived channel we need to compute the actions of 
$d_{0,\dn}$ and $d_{0,\up}^\dagger$ on $\bra{\chi_\sigma}$ and $\ket{\chi_\sigma}$:
\begin{align}
    \bra{\chi_\up} \, d_{0\dn} \, \frac{1}{\Egs-\mathcal{H}^{(0)}} \, d_{0\up}^\dagger \, \ket{\chi_\dn} &=
    -\frac{1}{\sqrt3} \,\bra{\up,\dn}_0 \, \bra{\Rt}_{\pm} \, \frac{1}{\Egs-\mathcal{H}^{(0)}} \,
    \frac{1}{\sqrt3} \, \ket{\up,\dn}_0\, \ket{\Rt}_{\pm}
    \label{eq:sign_change_2T}    
    \\
    \bra{\chi_\up}\,d_{0\up}^\dagger \, \frac{1}{\Egs-\mathcal{H}^{(0)}} \, d_{0\dn} \ket{\chi_\dn} 
    &= \frac{1}{\sqrt3} \, \bra{\Rt}_{\pm} \, \frac{1}{\Egs-\mathcal{H}^{(0)}} \, \frac{1}{\sqrt3} \, \ket{\Rt}_{\pm}
\end{align}
Assuming particle-hole symmetry, acting with $\mathcal{H}^{(0)}-\Egs$ on $\ket{\up,\dn}_0\otimes\ket{\Rt}_{\pm}$ 
and on $\ket{\Rt}_{\pm}$ yields the charging energy for adding/removing and electron to the 2T unit 
$\delta{E}_0=\mathcal{U}_0/2$, and therefore
\begin{align}
    J_0 = -2 \,\bar{V}_0^2 
    \, \left[ \left(-\frac{1}{3}\right) \left(-\frac{1}{\delta{E}_0} \right) 
    - \left(\frac{1}{3}\right)  \left(-\frac{1}{\delta{E}_0} \right)  \right]
    = -\frac{4}{3} \cdot \frac{\bar{V}_0^2}{\delta{E}_0} < 0
\end{align}
where the sign change of the first term in square brackets arises due to anticommutation 
of fermionic operators in Eq.~(\ref{eq:sign_change_2T}).
Hence the Kondo exchange coupling via the 2T unit, $J_0$, is ferromagnetic in nature. 

\subsubsection*{Kondo exchange due to coupling via 3T unit}

For the exchange coupling $J_\pm$ of the two 3T derived channels we now have to compute
the actions of $d_{\pm,\dn}$ and $d_{\pm,\up}^\dagger$ on $\bra{\chi_\sigma}$ and $\ket{\chi_\sigma}$,
which now yields two terms for each case:
\begin{align}
    \label{eq:3T:chi_up_d_dn}
    \bra{\chi_\up} \, d_{+\dn} &= 
        \tfrac{1}{\sqrt6}  \, \bra{0\up,+\up,+\dn,-\dn} 
        - \sqrt{\tfrac{2}{3}}\, \bra{0\dn,+\up,+\dn,-\up}  
    \\
    \label{eq:3T:d+_up_chi_dn}
    d_{+\up}^\dagger \ket{\chi_\dn} &= 
          -\tfrac{1}{\sqrt6}  \, \ket{0\dn,+\up,+\dn,-\up}
        +\sqrt{\tfrac{2}{3}}\, \ket{0\up,+\up,+\dn,-\dn}
    \\
    \label{eq:3T:chi_up_d+_up}
    \bra{\chi_\up} \, d_{+\up}^\dagger &= 
          -\tfrac{1}{\sqrt6}  \, \bra{0\up,-\dn}
          +\sqrt{\tfrac{2}{3}}\, \bra{0\dn,-\up}
    \\
    \label{eq:3T:d_dn_chi_up}
    d_{+\dn} \ket{\chi_\dn} &= 
        -\tfrac{1}{\sqrt6}  \, \ket{0\dn,-\up}
        +\sqrt{\tfrac{2}{3}}\, \ket{0\up,-\dn}
\end{align}
Assuming particle-hole symmetry, acting with $\mathcal{H}^{(0)}-\Egs$ on any of the states on the right hand sides of (\ref{eq:3T:chi_up_d_dn}-\ref{eq:3T:d_dn_chi_up}) now yields the charging energy for adding/removing an electron on the 
3T unit $\delta{E}_\pm=\frac{3}{4}\mathcal{U}_\pm$. Putting it all together, we obtain:
\begin{align}
    J_\alpha &= 
    -2\,\bar{V}_{\pm}^2\,\left(-\frac{1}{\delta{E}_\pm}\right) \left[ \frac{1}{3}+\frac{1}{3}
    -\left(-\frac{1}{3}\right)-\left(-\frac{1}{3}\right)\right] = \frac{8}{3}\cdot\frac{\bar{V}_{\pm}^2}{\delta{E}_\pm} > 0
\end{align}
where now the sign changes of the 3rd and 4th term in square brackets arise from anticommution
of fermionic operators in Eqs.~(\ref{eq:3T:chi_up_d_dn}-\ref{eq:3T:d_dn_chi_up}).
Thus the Kondo exchange couplings $J_\pm$ for the two 3T derived channels is antiferromagnetic 
in nature. 
In total this yields a three-channel Kondo model with mixed ferromagnetic and antiferromagnetic couplings.
More precisely, one ferromagnetic channel ($\alpha=0$) derived from coupling via the 2T unit, and two 
antiferromagnetic channels ($\alpha=\pm$) dervived from coupling via the two ZMs of the 3T unit.

\subsection*{Scaling theory for the M3CK model}

In this section we describe in more detail the scaling theory for three-channel Kondo model of 
mixed ferromagnetic and anti-ferromagnetic character given by Eqs.~(\ref{eq:H_K},\ref{eq:SW}).
It is based on Anderson's poor man's scaling approach for the Kondo model~\cite{Anderson:JPCSSP:1970}.
Subsequently, poor man's scaling has been used for understanding Kondo physics in more realistic situations of
multi-orbital magnetic impurities in metallic hosts, predicting underscreened and overscreened
Kondo effects~\cite{Nozieres:JP:1980}. More recently, poor man's scaling has been applied to understand the
observed narrowing of the Kondo resonance with the spin of magnetic impurities~\cite{Nevidomskyy:PRL:2009}.
An introduction and overview of the poor man's scaling approach can be found in Refs.~\cite{Hewson:book:1997,Nevidomskyy:chapter:2015}.

The basic idea of scaling theory is to successively integrate out the high-energy degrees of freedom, 
i.e. the conduction electrons at the band edges, in order to obtain an effective Hamiltonian 
$\HKT(\Lambda)$ valid on a lower energy scale given by the energy band cutoff $\Lambda$ for 
the conduction electrons.
In the case of Kondo-type models the effective Hamiltonian 
has the same form as the original Hamiltonian, but with renormalized interactions 
$\JKT{\alpha}$:
\begin{equation}
    \HKT(\Lambda) = \HK[\{\tilde{J}_\alpha(\Lambda)\}].
\end{equation}
One then follows the flow of the effective Hamiltonian $\HKT(\Lambda)$, 
as the energy cutoff $\Lambda$ decreases to a fixed point that describes 
the low-energy excitations of the system. 

Scaling theory leads to a set of coupled differential equations called 
scaling equations which describe the change of the exchange couplings, 
as the band cutoff $\Lambda$ is reduced. In the poor man's scaling approach to scaling theory 
the coupling to the high-energy states is taken into account by perturbation theory to finite order.
It is straight forward to generalize the scaling equations to third order for the
multi-channel Kondo model with all exchange couplings equal 
given in Ref.~\cite{Nozieres:JP:1980} to a multi-channel Kondo model 
with different couplings for each channel:
\begin{align}
  \label{eq:scaling1}
  \Lambda\,\frac{d\tilde{J}_\alpha}{|d\Lambda|} 
  &= 2 \,\rho_{\rm c} \, \tilde{J}_\alpha^2 - 2\sum_{\alpha^\prime} (\rho_{\rm c}\,\tilde{J}_{\alpha^\prime})^2 \, \tilde{J}_\alpha
\end{align}
where $\rho_{\rm c}$ is the conduction electron density of states, which we assume to be equal and constant 
for all three channels $\alpha=0,+,-$. 
Multiplying the scaling equations by $\rho_{\rm c}$, defining the dimensionless exchange couplings 
$\CJ_\alpha\equiv\rho_{\rm c}\,\tilde{J}_\alpha$, and using $|d\Lambda|/\Lambda=-d\ln\Lambda$
leads to the scaling equations (\ref{eq:scaling}) in the main text.

The first term on the r.h.s. of (\ref{eq:scaling1})
is the second-order contribution which only describes the renormalization within each channel.
Thus, to second order the scaling equations only would describe three independent one-channel 
Kondo models. Note that the second order contribution is always positive. Thus for initial ferromagnetic 
coupling ($J_\alpha<0$) in leading order $J_\alpha$ always decreases in magnitude,
eventually going to zero, as the band cutoff goes to zero.
The third-order contribution, on the other hand, given by the second term on the r.h.s. of (\ref{eq:scaling1})
describes coupling between the channels, giving rise to deviations from the purely single-channel behavior 
described by second order. 

The scaling equations are a system of ordinary first order differential equations, also known in mathematics 
as an autonomous system. We solve the scaling equations (\ref{eq:scaling}) with the ordinary differential 
equation solver \verb|solve_ivp| implemented in the Python package \verb|SciPy| using the Runge-Kutta algorithm. Fig.~\ref{fig:theory}(d) shows the solutions of the autonomous system (\ref{eq:scaling}), 
also called scaling trajectories in this context. The scaling trajectories show the flow of the exchange
couplings $\{\JKT\alpha\}$ as the high-energy degrees of freedom are eliminated successively, 
i.e., the energy cutoff $\Lambda$ is reduced, corresponding to reducing temperature and/or bias voltage in an 
STS experiment.

The initial conditions for the scaling equations are set by the exchange couplings obtained from the Schrieffer--Wolff transformation, Eq.~(\ref{eq:SW}). Expressed in terms of the dimensionless couplings \( \CJ_\alpha \), the initial conditions read:
\begin{align}
\label{eq:initial-condition}
    \CJ_0 = -\frac{4}{3}\frac{\bar{V}_0^2\rho_c}{\delta{E}_0} = -\frac{4}{3}\frac{\Gamma_0/\pi}{\delta{E}_0} \sim -0.07
    \hspace{1ex}\mbox{ and }\hspace{1ex}
    \CJ_\pm = \frac{8}{3}\frac{\bar{V}_\pm^2\rho_c}{\delta{E}_\pm} = \frac{8}{3}\frac{\Gamma_\pm/\pi}{\delta{E}_\pm}
    \sim 0.075
\end{align}
where we have used the hybridization strengths $\Gamma_0\sim55\,\meV$ and $\Gamma_\pm\sim35\,\meV$, 
as well as the charging energies $\delta{E}_0\sim334\,\meV$ and $\delta{E}_\pm\sim394\,\meV$, 
see Eq.~(\ref{eq:charging-energies}) in Methods.

Fig.~\ref{fig:theory}d shows representative scaling trajectories of the exchange couplings $\CJ_0(\Lambda)$ and $\CJ_\pm(\Lambda)$ obtained from Eq.~(\ref{eq:scaling}) for different initial conditions $\CJ_0(\Lambda_0)$ and $\CJ_\pm(\Lambda_0)$.
Arrows indicate the renormalization flow as the cutoff $\Lambda$ is reduced, placed at 
equal time steps $t=-\ln\Lambda/\Lambda_0$, each corresponding to a reduction of the cutoff by a factor of $e^{-1}$. 
We first consider the limiting cases of purely ferromagnetic coupling via the 2T unit (red trajectory) and purely antiferromagnetic coupling via the 3T unit (blue trajectories), before turning to the general case in which both couplings are present (gray trajectories).

For combined ferromagnetic coupling via the 2T unit, $\CJ_0(\Lambda_0)<0$, 
and antiferromagnetic coupling via the 3T unit, $\CJ_\pm(\Lambda_0)>0$, all scaling trajectories 
(gray) flow to the fixed point of the M3CK model, $\CJ^\ast_{\rm M3CK} = (0,1/2)$, consisting of the weak coupling fixed point for the 2T channel and the two-channel fixed point 
    for the 3T channels. 
For the experimentally relevant initial conditions of the 2T--3T dimer on Au(111), given by Eq.~\ref{eq:initial-condition} (purple circle in Fig.~\ref{fig:theory}d), the resulting scaling trajectory is shown in purple in 
Fig.~\ref{fig:theory}d. The renormalization flow initially remains in the vicinity of the weak coupling
fixed point associated with the ferromagnetic Kondo regime, exhibiting the characteristic logarithmically slow evolution reflected by the increasing arrow density. At cutoff values around $1.5\,\meV$ (orange filled circle) corresponding to an experimental temperature $T\sim4.6\,$K and lock-in modulation $V\sim1\,\meV$ (assuming that temperature and lock-in broadening
are approximately additive~\cite{Turco:PRR:2024}),
the system is therefore still governed by the weak coupling fixed point. At lower energy scales, the trajectory crosses over towards the M3CK fixed point, where the flow accelerates and approaches a constant rate, driven by the growing—yet bounded—antiferromagnetic couplings of the two 3T channels. 
Notably, the 2T channel remains ferromagnetic ($\CJ_0 < 0$) throughout the flow and vanishes only upon reaching the M3CK fixed point. 

Owing to the finite and small energy scale of the two-channel fixed point,  
overscreening on the 3T unit actually facilitates the emergence of the ferromagnetic 
Kondo effect on the 2T unit. Hence the antiferromagnetic exchange coupling 
does not have to be suppressed in order to observe the ferromagnetic Kondo 
effect, contrary to a model with only one antiferromagnetic channel~\cite{ternes_probing_2017}. 
In fact here both couplings are of the same order: $|\CJ_0|\sim|\CJ_\pm|$ (see Eq.~\ref{eq:initial-condition}).

\subsection*{Experimental details}

STM and STS data presented in the manuscript were performed with two distinct low-temperature STM setups. Data reported Fig. \ref{fig:intro} measurements were performed with a commercial low-temperature STM from Scienta Omicron operated at a temperature of $4.5$ K and a base pressure below $5 \cdot 10^{-11}$ mbar. The corresponding differential conductance \didv spectra were obtained with a lock-in amplifier (Zurich Instruments). Modulation voltages for each measurement are provided in the respective figure caption. Data of Fig. \ref{fig:B-field IETS} was recorded at a temperature of 54 mK using a home-built dilution refrigerated UHV STM \cite{balashov_compact_2018}. \didv spectra were recorded using a lock-in amplifier at a modulation frequency of 3.2 kHz and modulation amplitude of tens of $\mu\eV$ (noted in captions). An energy resolution of the instrument of 9.3 $\mu\eV$ was estimated from the Josephson peak in tunneling experiments between a superconducting tip and sample.
All bias voltages are reported with respect to the sample. Unless otherwise stated, metallic tips were used for all measurements. Data analysis was performed using the Igor Pro software package (Wavemetrics). 

\subsubsection*{Sample preparation}
Au(111) single-crystal surfaces were cleaned by repeated cycles of Ar$^+$ sputtering and annealing. The cleanliness and quality of the surfaces were confirmed by STM imaging prior to molecular deposition. For comprehensive details on the in-solution and on-surface synthesis of the 2T–3T dimer, along with its structural and electronic characterization, we refer the reader to Ref.~\cite{turco_arxiv}.

\backmatter

\section*{Data availability}
The main data supporting the ﬁndings of this study are available within
the article and its Supplementary Information. Source data of Figs. 1,
2 and 3 are provided with this paper. Extra data are available from the
corresponding authors upon request.

\bibliography{refs}

\section*{Acknowledgements}
We would like to thank Joaquín Fernández-Rossier, João C.~G. Henriques and Gon\c{c}alo Catarina for insightful scientific discussions.

\section*{Author contributions}
ET, NK, HC and SG performed the measurements, while de experiments were designed by ET, WW, RF, PR and DJ. 
The analysis of experimental data was done by ET and NK, whereas DJ provided theoretical description. 
All authors contributed to the writing of the manuscript.

\section*{Funding}
This research was financially supported by the EU Graphene Flagship (Graphene Core 3, 881603), H2020-MSCA-ITN (ULTIMATE, No. 813036), Swiss National Science Foundation (SNF-PiMag, No. CRSII5\_205987 and 212875), the Werner Siemens Foundation (CarboQuant), the Deutsche Forschungsgemeinschaft (DFG, German Research Foundation) through the Collaborative Research Center ”4f for Future” (CRC 1573, project number 471424360) project B2 and DFG project Wu 349/17-1, and the “Plan Gen-T of Excellence” of Generalitat Valenciana through Grant No. CIDEXG/2023/7.

\section*{Competing interests}
The authors declare no competing interests.

\section*{Additional information}
\bmhead{Supplementary Information} The online version contains
supplementary material available at (weblink), with additional 
computational results, additional experimental data, an in-depth 
analysis of the magnetic-field-dependent measurements, and further 
details on the data processing and analysis procedures.

\section*{Figure Legends/Captions (for main text figures)}
    
\bmhead{Figure 1}
\textbf{Magnetic characterization of 2T-3T dimer}.
\textbf{a} STM image of an isolated 2T-3T dimer on Au(111) ($V = -0.1$ V, $I = 100 $pA, scale bar = 0.5 nm). \textbf{b,c} Chemical structure and Heisenberg representation for the 2T-3T dimer. \textbf{d} Representation of the three zero modes, where the circle size indicates the absolute value of the wave function, and the arrows encode the complex phase. \textbf{e} Magnetic spectrum derived from the analytical solution of the Heisenberg dimer model for an antiferromagnetic coupling between the 2T and 3T units. \textbf{f} Low-bias STS spectra acquired with a carbon monoxide functionalized tip, on two distinct locations on the molecule, indicated in Fig.~\ref{fig:intro}a with colored filled circles. The reference spectrum, taken with the same tip on bare Au(111), is offset by -6\,nS (grey line). 
    Open feedback parameters \didv spectra: $V=-80$ mV, $I=850$ pA; Lock-in modulation $V_{\rm m}=1 $ mV; $T_\text{sample}=4.5$ K. \textbf{g} Orbital-resolved spectral function calculated with the one-crossing approximation at $T=4.6$ K for particle-hole symmetry and the interactions given by (\ref{eq:interactions}). The red and blue colors denote the 2T and 3T units, respectively.
 
\bmhead{Figure 2}
\textbf{Theoretical description of the Kondo effect in the 2T-3T dimer}.
\textbf{a,b} Illustrations of the two models employed here for the description of 2T-3T dimer 
coupled to conduction electrons:
\textbf{a} Three-orbital Anderson impurity model of 2T-3T dimer
coupled to three independent conduction electron baths, Eqs.~(\ref{eq:H_imp},\ref{eq:H_AIM}); \textbf{b} effective 
Kondo model of spin-1/2 with three screening channels, Eqs.~(\ref{eq:H_K},\ref{eq:SW}).
\textbf{c} Illustrations of the exchange processes between the dimer and substrate conduction electrons mediated by the 2T and 3T units, illustrating ferromagnetic and antiferromagnetic coupling mechanisms.
\textbf{d} Third-order renormalization-group scaling trajectories of the exchange couplings $\CJ_0$ and $\CJ_\pm$ for the M3CK model, Eq.~(\ref{eq:scaling}).
The red trajectory corresponds to a  ferromagnetic one-channel Kondo model, 
and the blue trajectory to an antiferromagnetic two-channel Kondo model. 
The purple trajectory denotes the physical scaling flow relevant to the 2T--3T dimer, 
determined by the initial conditions in Eq. (\ref{eq:initial-condition}) (purple circle).
The yellow star marks the weak coupling fixed point of the ferromagnetic Kondo effect, while 
light blue star indicates the M3CK fixed point, characterized by the coexistence of 
ferromagnetic Kondo effect in the 2T channel and an overscreened Kondo effect in the 3T channels. The orange filled circle denotes the cutoff corresponding to the experimental situation ($T_{\rm sample}\sim4.5$K and $V_{\rm m}\sim1$mV).

\bmhead{Figure 3}
\textbf{Magnetic-field-dependent splitting of the overscreened Kondo peak and ferromagnetic Kondo dip.} 
\textbf{a,c} Series of \didv spectra (solid black line) acquired with a metal tip at 3\,T and 2\,T units, respectively, as a function of the applied magnetic field $B_0$. All spectra are normalized, to correct for dynamical Coulomb blockade (details in SI). Theoretical fit functions, using a third-order perturbative scattering model~\cite{Ternes:NJP:2015} with Frota lineshapes convolved with the Fermi–Dirac distribution to account for thermal broadening, are shown as solid lines: blue for 3T and red for 2T.
The spectra for $B_0=$ 0\,T, 1\,T, 2\,T, 3\,T, 4\,T are offset for clarity by 80\,nS, 50\,nS, 24\,nS, 32\,nS, 25\,nS in panel a and by 8\,nS, 7.5\,nS, 6\,nS, 2.3\,nS, 7\,nS in panel c. The raw data without Coulomb blockade corrections and offsets are shown in Fig.~S5.
\textbf{b} Effective magnetic field $B_\mathrm{fit}$ obtained from the fits to the magnetic field evolution of the overscreened Kondo peak (assuming $g=2$), plotted as a function of the applied magnetic field $B_0$. These values of $B$ are then held fixed when fitting the magnetic field dependence of the ferromagnetic Kondo dip shown in \textbf{c}. 
\textbf{d} Renormalized $g$-factor as a function of the applied magnetic field $B_0$, calculated for different values of Kondo exchange coupling $\CJ_\pm$, according to Eq.~(\ref{eq:renorm_g-factor}). The experimental values $g_{\mathrm{exp}}$, extracted from \textbf{b}, are shown as black filled circles with their error bars. The best fit ($\CJ_\pm \approx 0.62$) is shown as a solid red line, while the red dashed line indicates the $g$-factor corresponding to $\CJ_\pm \approx 0.75$, obtained via a Schrieffer--Wolff transformation. The fit was performed by $\chi^2$ minimization using the experimental uncertainties, yielding a reduced $\chi^2$ of 0.23. Error bars in panels b,d correspond to $\pm \sigma$. Open-feedback parameters: $V = -20$\,mV, $I = 1$\,nA; lock-in modulation $V_{\rm m} = 80$\,$\mu$V; $T_\text{sample} = 54$\,mK.

\end{document}


\title{Supplemental Material:\\ Observation of the Ferromagnetic Kondo Effect}

\author{Elia Turco}
\email{e.turco@tudelft.nl}
\affiliation{nanotech@surfaces Laboratory, Empa - Swiss Federal Laboratories for Materials Science and Technology, 8600 Dübendorf, Switzerland}
\altaffiliation{Current address: QuTech and Kavli Institute of Nanoscience, Delft University of Technology, 2600 GA Delft, The Netherlands}

\author{Nils Krane}
\affiliation{nanotech@surfaces Laboratory, Empa - Swiss Federal Laboratories for Materials Science and Technology, 8600 Dübendorf, Switzerland}


\author{Hongyan Chen}
\affiliation{Physikalisches Institut, Karlsruhe Institute of Technology, 76131, Karlsruhe, Germany}

\author{Simon Gerber}
\affiliation{Physikalisches Institut, Karlsruhe Institute of Technology, 76131, Karlsruhe, Germany}

\author{Wulf~Wulfhekel}
\affiliation{Physikalisches Institut, Karlsruhe Institute of Technology, 76131, Karlsruhe, Germany}

\author{Roman Fasel}
\affiliation{nanotech@surfaces Laboratory, Empa - Swiss Federal Laboratories for Materials Science and Technology, 8600 Dübendorf, Switzerland}
\affiliation{Department of Chemistry, Biochemistry and Pharmaceutical Sciences, University of Bern, 3012 Bern, Switzerland}

\author{Pascal Ruffieux}
\affiliation{nanotech@surfaces Laboratory, Empa - Swiss Federal Laboratories for Materials Science and Technology, 8600 Dübendorf, Switzerland}

\author{David Jacob}
\email{david.jacob@ua.es}
\affiliation{Departamento de Física, Universidad de Alicante, Campus de San Vicente del Raspeig, E-03690 Alicante, Spain}

\maketitle

\renewcommand{\theequation}{S\arabic{equation}}

\section{Hubbard model for 2,3-dimer}
\label{sec:hubbard}

It is generally established that nanographenes are well described by a Hubbard-type Hamiltonian
that also includes third-neighbor hopping~\cite{Ortiz:NL:2019,Jacob:PRB:2022}:
\begin{equation}
\label{eq:Hubbard}
  \mathcal{H} = -t_1 \sum_{\langle i,j \rangle, \sigma} c_{i\sigma}^\dagger c_{j\sigma} - t_3 \sum_{\langle\langle\langle i,j \rangle\rangle\rangle, \sigma} c_{i\sigma}^\dagger c_{j\sigma}
  + U \sum_i \hat{n}_{i\up} \hat{n}_{i\dn}
\end{equation}
where $t_1$ is the hopping between nearest-neighbor sites, $t_3$ is the third-neighbor hopping,
and $U$ is the on-site Coulomb repulsion. For future reference we will denote the interaction 
term of the Hubbard model by $\mathcal{H}_U$.
In this description each site of the Hubbard model represents a $p_z$-orbital of a carbon atom.

\begin{figure}
  \begin{center}
    \includegraphics[width=\linewidth]{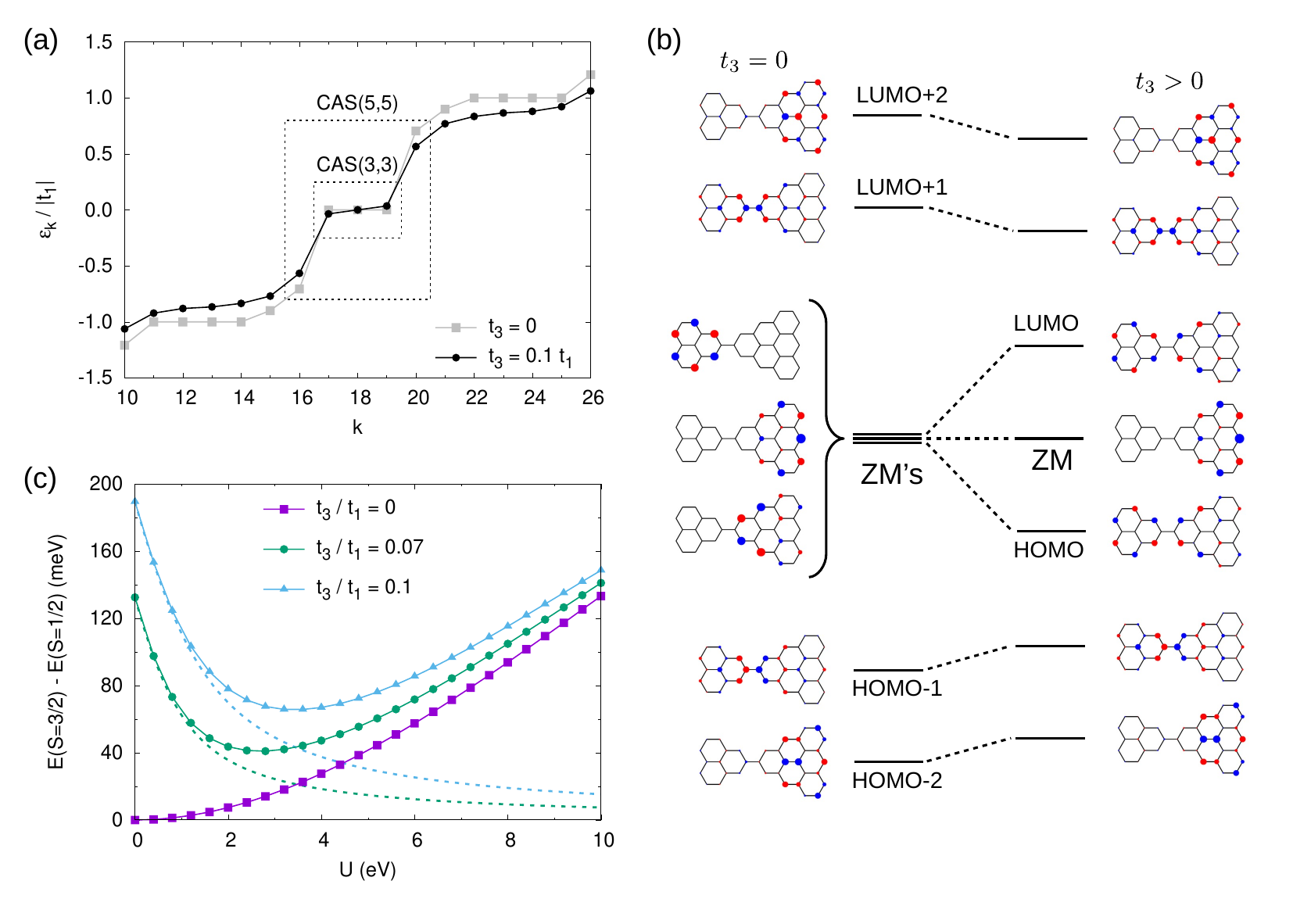}
    \end{center}
  \caption{\label{fig:hubbard} Results for Hubbard model (\ref{eq:Hubbard}) of 2,3-dimer for $t_1=2.7\,\eV$. 
  (a) Single-particle energies ($U=0$) for $t_3=0$ (grey squares) and for $t_3>0$ (black circles).
  (b) Molecular orbitals (also $U=0$) close to zero energy for $t_3=0$ (left) and for $t_3>0$ (right).
  (c) Energy splitting of the first excited state with $S=3/2$ from the GS with $S=1/2$
  of the 2,3-dimer as a function of the Hubbard $U$ for a CAS(5,5) consisting of the 
  the 3 ZM's and the HOMO-1 and LUMO+1 for different values of $t_3/t_1$ as indicated.
  The dashed lines show the corresponding energy splitting for CAS(3,3), see text for details.
  }
\end{figure}

Fig.~\ref{fig:hubbard}(a) shows the single-particle spectrum ($U=0$) of the Hubbard model corresponding
to the 2,3-dimer for $t_1=2.7\,\eV$ and for both $t_3=0$ and $t_3>0$. In the absence of third-neighbor hopping there
are three zero-energy modes or for short zero modes (ZMs), corresponding to the ZMs of the
uncoupled triangulene species. As shown in Fig.~\ref{fig:hubbard}(b) the ZMs can be localized on
the invidual triangulenes (one on the 2T and two on the 3T). When third-neighbor hopping is
switched on ($t_3>0$), one of the ZMs of the 3T unit couples to the ZM of the 2T unit, leading
to an energy splitting $\delta\propto\,t_3$. The other ZM on the 3T unit does not couple to the 2T unit
due to the lack weight at the carbon sites connected to the 2T unit. Therefore this ZM remains at zero
energy even in the presence of $t_3$.

Similar to previous work~\cite{Jacob:PRB:2021,Jacob:PRB:2022}
we now solve the many-body problem of the interacting Hubbard model ($U>0$) for the 2,3-dimer
by exact diagonalization of the complete active space (CAS) consisting of the ZMs plus the
molecular orbitals closest in energy to the ZMs as indicated by the boxes in Fig.~\ref{fig:hubbard}(a).
The results are shown in Fig.~\ref{fig:hubbard}(c).
For $t_3=0$ and only taking into account the three ZMs in CAS(3,3) there is no coupling between
the 2T and 3T units and thus no exchange splitting (not shown). Hence the ground state (GS)
is a sextuplett formed by the $S=1/2$ doublet of the 2T unit and the $S=1$ triplet of the 3T unit.
When $t_3$ is switched on, the coupling between the 2T and 3T ZMs leads to antiferromagnetic
kinetic exchange $J_{\rm kin}\sim4|t_3|^2/U$~\cite{Jacob:PRB:2022}. 
Thus the GS becomes a total $S=1/2$ doublet GS,
split from the excited $S=3/2$ quadruplet by the exchange energy $J_{\rm kin}\sim4|t_3|^2/U$,
as shown by the dashed light-blue and purple curves in Fig.~\ref{fig:hubbard}(c). 

Taking into account single-particle states beyond CAS(3,3) gives rise to so-called
Coulomb-driven superexchange $J_{\rm coul}$ via intermediate states which have weight in
both triangulene units~\cite{Jacob:PRB:2022}. Due to the direct coupling between the two
triangulene unit, this superexchange mechanism is antiferromagnetic in nature, as discussed
in Ref.~\onlinecite{Jacob:PRB:2022}, and leads to additional splitting between the 
$S=1/2$ GS and the $S=3/2$ excited state, quadratic in $U$ for not too large values of $U$.
This is demonstrated by the purple curve with squares in Fig.~\ref{fig:hubbard}(c),
which shows a CAS(5,5) calculation in the absence of third-neighbor hopping ($t_3=0$)
and thus in the absence of kinetic exchange coupling. When $t_3$ is switched on both
kinetic exchange $\sim1/U$ and Coulomb-driven superexchange $\sim{U^2}$ contribute,
giving rise to a minimum in the splitting between the $S=1/2$ GS and the $S=3/2$ excited states,
as shown by the curves with green circles and light-blue triangles in Fig.~\ref{fig:hubbard}(c). 
On the other hand, also taking into account the HOMO-2 and LUMO+2 states in a CAS(7,7) calculation
does not lead to additional splitting (not shown), as these states cannot mediate the Coulomb-driven
superexchange due to their localization in the 3T unit, see Fig.~\ref{fig:hubbard}(b).

For modeling the 2T-3T dimer on the Au surface we choose Hubbard model parameters $U=4\,\eV$, $t_1=2.7\,\eV$ and $t_3=0.07\,t_1$ which yields a spin excitation energy of $\sim47\,\meV$. This is slightly larger than the experimentally observed one of $\sim40\,\meV$, as we expect the exchange coupling to the conduction electrons in the Au substrate to renormalize the spin excitation energy to a somewhat lower value, as discussed in previous works~\cite{Oberg:NNano:2014,Jacob:PRB:2021,Krane:NL:2023}.

\section{Coupling of the zero-modes to the substrate from density functional theory}
\label{sec:Hybridization}

\begin{figure}[h]
  \includegraphics[width=0.8\linewidth]{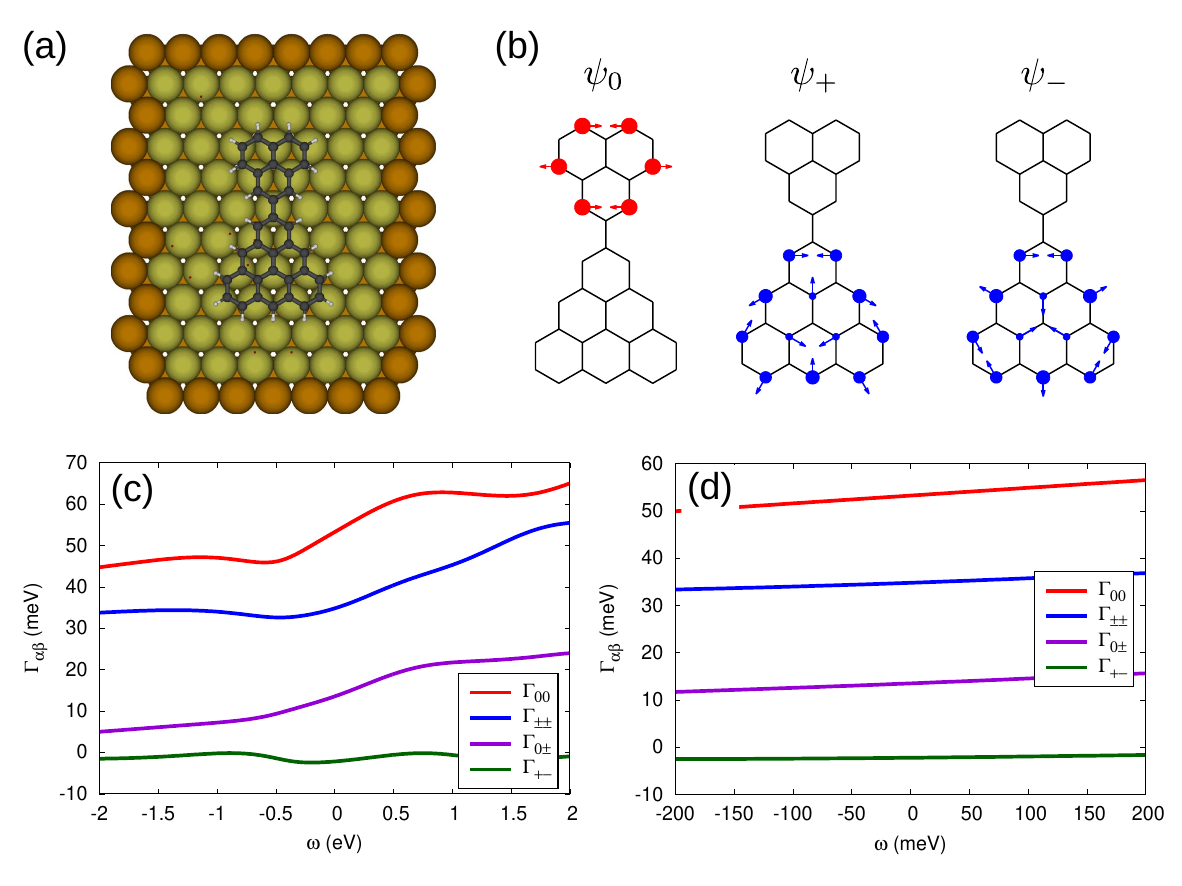}
  \caption{\label{fig:hybfunc}
    (a) Atomic structure of 2T-3T molecule on Au(111) surface.
    (b) $C_3$-symmetric zero-modes of the 2T ($\psi_0$) and 3T ($\psi_+,\psi_-$) units.
    (c) Energy-dependent broadening $\Gamma_{\alpha\beta}(\omega)=-\Im\,\Delta_{\alpha\beta}(\omega)$
    of the 2T zero-mode $\psi_0$ and the 3T zero-modes ($\psi_+,\psi_-$) calculated 
    within DFT in the GGA using the quantum transport code ANT.G~\cite{Jacob:JCP:2011}.
    (d) Same as (c) but for a smaller energy window ($|\omega|\le200\,\meV$).
  }
\end{figure}

In order to determine how the zero-modes of the 2T-3T molecule couple
to the conduction electrons in the substrate, we have
performed density functional theory (DFT) calculations for the 2T-3T molecule
on the Au(111) surface, as shown in panel (a) of Fig.~\ref{fig:hybfunc}.
To this end we have employed the ANT.G {\it ab initio} quantum transport code~\cite{Jacob:JCP:2011}.
In a first step the electronic structure of the 3T molecule on a finite
piece of the Au(111) surface (yellow atoms), embedded into an effective
model described by Bethe lattice atoms (brown atoms) representing the
rest of the surface, is computed on the level DFT. 
We are not so much interested in an accurate quantitative description
of the coupling of the orbitals to the substrate, but rather in a 
qualitative evaluation of the overall screening situation as an 
input for our Anderson impurity and Kondo models.
In particular, we want to determine, whether the zero-modes couple to 
each other via the substrate. Secondly, we also want to obtain a rough 
estimate of the coupling strengths.

Therefore we consider an idealized situation where the molecule
is in the ideal flat graphene geometry (1.42\r{A} nearest-neighbor distance) 
lying 3.2\r{A} above an ideal Au(111) surface (2.885\r{A} nearest-neighbor distance). 
We employ a minimal basis set together with the generalized gradient
approximation (GGA) in the parametrization due to Perdew, Burke and Ernzerhof~\cite{Perdew:PRL:96}. In a second step the so-called hybridization 
function matrix ${\bm\Delta}(\omega)\equiv(\Delta_{\alpha\beta}(\omega))$ 
projected onto the three zero-modes of the 2T-3T molecule shown in panel (b)
is calulated from the Kohn-Sham Green's function of the embedded
molecule on the surface, as described in Ref.~\cite{Jacob:JPCM:2015}.
The hybridization function describes coupling of the molecular orbitals to the
conduction electrons in the substrate. Specifically, the negative imaginary part
$\bm{\Gamma}(\omega)=-\Im\bm\Delta(\omega)$ yields the lifetime broadening
of the molecular orbitals due to the coupling to the substrate. Panels (c,d)
of Fig.~\ref{fig:hybfunc} show $\bm{\Gamma}(\omega)$ for the two ZMs of the
2T-3T molecule. As expected, both zero-modes of the 3T unit couple equally to the substrate, 
$\Gamma_{++}=\Gamma_{--}\sim35\,\meV$ around the Fermi level, 
while the coupling of the 2T zero-mode is somewhat larger, $\Gamma_0\sim55\,\meV$ 
around the Fermi level. These values are similar to the ones
calculated for the 2T-2T dimer in Ref.~\cite{Krane:NL:2023}.
Importantly, the coupling between both 3T zero-modes via the substrate is negligible, 
i.e., $\Gamma_{+-}=\Gamma_{-+}\sim0$. Therefore the coupling of the
two zero-modes on the 3T unit to the substrate yields two \emph{independent} 
screening channels. On the other hand, the coupling between the 2T zero-mode 
($\psi_0$) and the two 3T zero-modes ($\psi_+,\psi_-$) is not negligible, 
$\Gamma_{0\pm}\sim14\,\meV$ around the Fermi level. However, in our effective 3-orbital
model (Eq.~1 in main text) we absorb it into the effective exchange coupling 
$J_{\rm eff}$ between the 2T and 3T units.

\section{Magnetic-field renormalization in the one-channel Kondo model}

For the one-channel Kondo model, the scaling equation for the exchange coupling 
to second order is given by~\cite{Hewson:book:1997}
\begin{equation}
    d\mathcal{J}/d\ln\Lambda = -2\,\mathcal{J}^2 ,
\end{equation}
which can be integrated analytically, resulting in:
\begin{equation}
     \mathcal{J}(\Lambda) = 1/(2\,\ln(\Lambda/\Lambda_0)+1/\mathcal{J}_0)
\end{equation}
where $\mathcal{J}_0\equiv\mathcal{J}(\Lambda_0)$ is the bare exchange coupling.
Plugging this result into the equation for the renormalized $g$-factor~\cite{Garst:PRB:2005} 
(see also Eq.~(6) in main text) and integrating, yields
\begin{equation}
\label{eq:gtilde}
    g_{\rm eff}(\Lambda) = g\,\exp\left(\mathcal{J}_0-\frac{1}{1/\mathcal{J}_0+2\,\ln(\Lambda/\Lambda_0)}\right)
\end{equation}
where $g$ is the Knight-shifted bare $g$-factor, given by $g=g_0-\mathcal{J}$ with $g_0=2$
the bare $g$-factor (see Ref.~\cite{Garst:PRB:2005}).
Taking the spin excitation energy to the spin-quadruplet as the initial cutoff, $\Lambda_0=40\,\meV$,
and assuming that the running cutoff $\Lambda$ is dominated by the magnetic field (see main text),
$\Lambda\sim g\,\mu_B\,B_0$, we obtain the renormalized $g$-factor as a function of the applied
magnetic field for the one-channel Kondo model.

\begin{figure}[ht]
  \includegraphics[width=\linewidth]{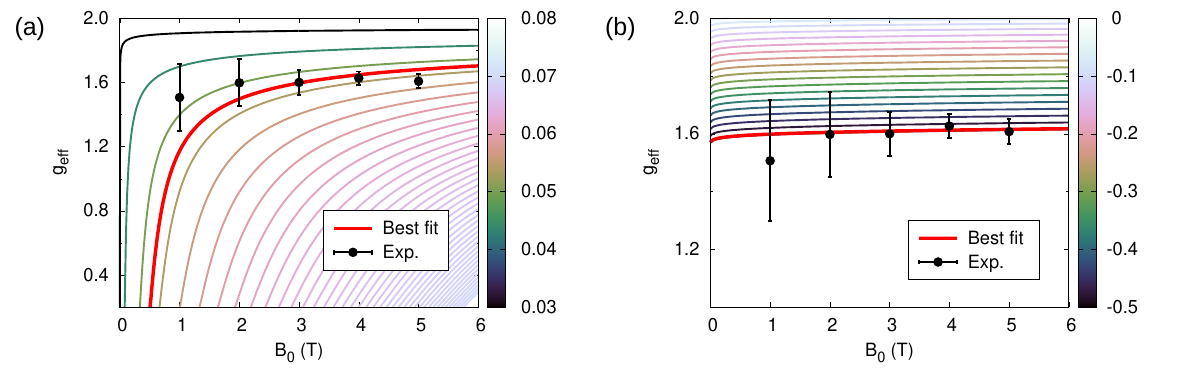}
  \caption{\label{fig:1CK-g}
    (a,b) Renormalized $g$-factor as a function of the applied magnetic field $B_0$
    for the one-channel Kondo model computed by scaling theory to second order
    for (a) anti-ferromagnetic exchange coupling ($\mathcal{J}>0$) and 
    (b) ferromagnetic exchange coupling ($\mathcal{J}<0$). 
    Full black circles show the effective $g$ extracted from experiment (see main text),
    and solid red lines show the best fits of $\mathcal{J}$ to the experimental data, i.e.,
    $\mathcal{J}\sim0.052$ (reduced $\chi^2\sim1.55$) for (a) and $\mathcal{J}\sim-0.498$ (reduced $\chi^2\sim0.09$) for (b).
  }
\end{figure}

Fig.~\ref{fig:1CK-g} shows the renormalized $g_{\rm eff}$ as a function of the magnetic field $B_0$ for
(a) antiferromagnetic coupling ($\mathcal{J}>0$) and (b) ferromagnetic coupling ($\mathcal{J}<0$)
in comparison with the experimental $g$-factor as a function of the applied magnetic field. 

For the antiferromagnetic case (Fig.~\ref{fig:1CK-g}a), the one-channel Kondo model yields a best-fit value $\mathcal{J}\sim0.05$, close to $\mathcal{J}\sim0.075$ obtained from the Schrieffer--Wolff transformation. However, the overall agreement with experiment remains limited. In particular, the field dependence $g_{\rm eff}(B)$ is significantly less well described than in the M3CK model: the reduced $\chi^2$ is $\sim1.55$ for the one-channel model, compared to $\sim0.23$ for the M3CK model (Fig.~3d). This substantial discrepancy demonstrates that the one-channel description fails to capture the observed behavior, whereas the M3CK model provides a quantitatively accurate account.
The reason that the one-channel Kondo model does not fit the data so well lies in 
the strong renormalization as the strong coupling fixed point
is approached ($\mathcal{J}\to\infty$ as $\Lambda\sim{g\,\mu_B\,B_0}\to0$), 
i.e., $g_{\rm eff}\approx0$ in the Kondo regime, when $\Lambda\sim g\,\mu_B\,B_0<k\,\TK$.

On the other hand, Fig.~\ref{fig:1CK-g}b shows that for the ferromagnetic case the 
renormalization of the $g$-factor is very weak, a consequence of the weak coupling
fixed point (where $\mathcal{J}\rightarrow0$) on the on hand, and of the Knight shift 
partially compensating the weak second order renormalization on the other hand. 
As a result, an unreasonably large value for the exchange coupling $\mathcal{J}\sim-0.5$ is
needed to fit $g_{\rm eff}(B_0)$ to the experimental data.

Overall, these results show that neither with ferromagnetic or 
antiferromagnetic coupling can the one-channel Kondo model 
adequately describe the experimentally measured $g$-factor.


\section{Correction of Conductance Spectra for DCB Dip}
At very low STM temperatures the emergence of the dynamic Coulomb blockade (DCB) can observed in $\mathrm{d}I/\mathrm{d}V$ spectra as a dip in conductance at zero bias voltage~\cite{Senkpiel:PRL:2020,Esat:PRB:2015, Ast:NatCom:2016}
.
We corrected the experimental data based on the the procedure as described in \cite{Esat:PRR:2023}.
The spectra were divided by a reference spectrum taken on Au(111) and then multiplied again with the mean of the reference spectrum.

\begin{figure}[ht!]
    \centering
    \includegraphics[width=0.6\linewidth]{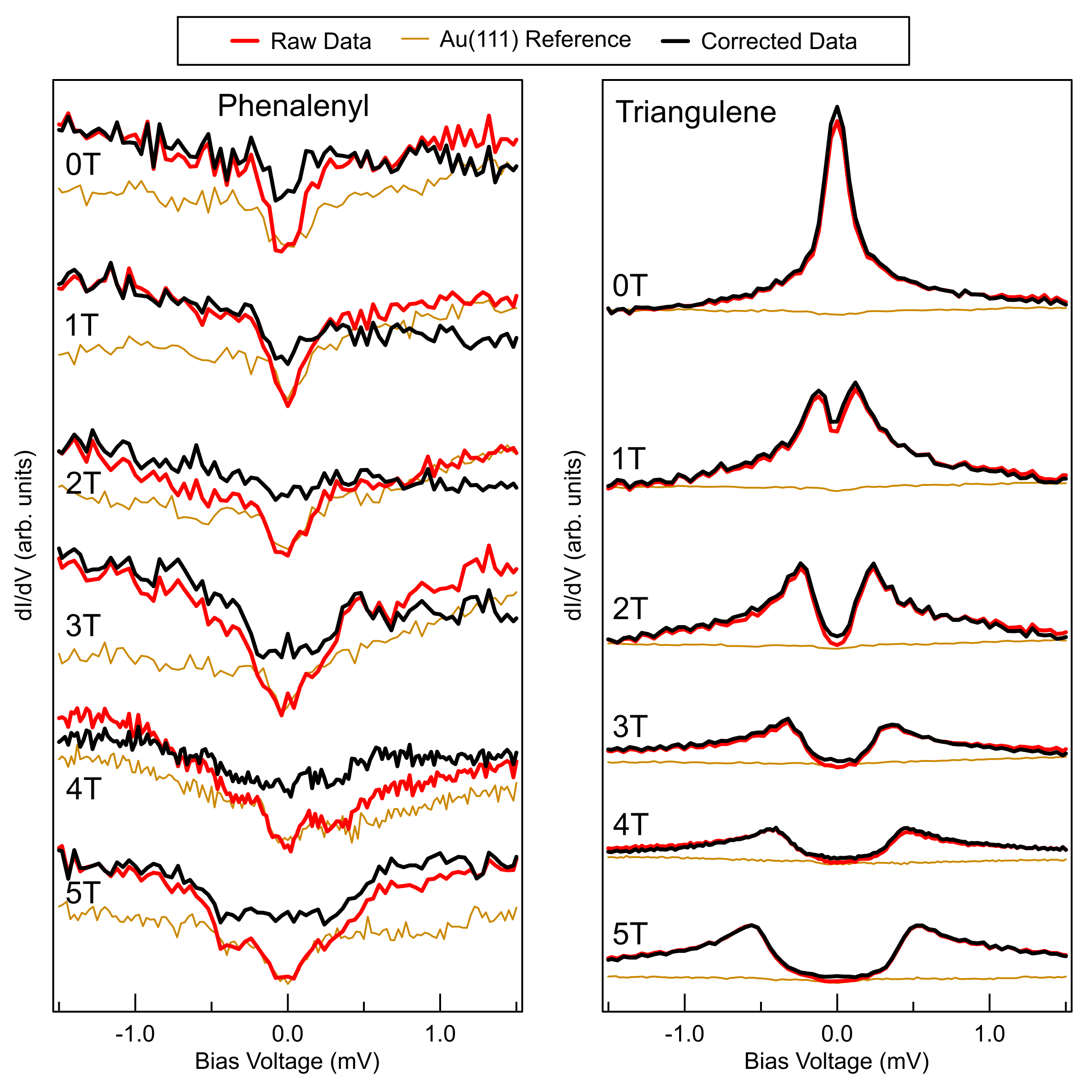}
    \caption{
    Correction of spectra for the DCB dip by division of the raw data (red) with a reference spectrum taken on Au(111) (gold). The spectra are shifted for clarity.
    }
    \label{fig:renormalization_DCB}
\end{figure}

\section{Additional data}

\subsection*{\didv analysis with same CO tip on 2T-H3T and 2T-3T}

\begin{figure}[ht!]
    \centering
    \includegraphics[width=0.6\linewidth]{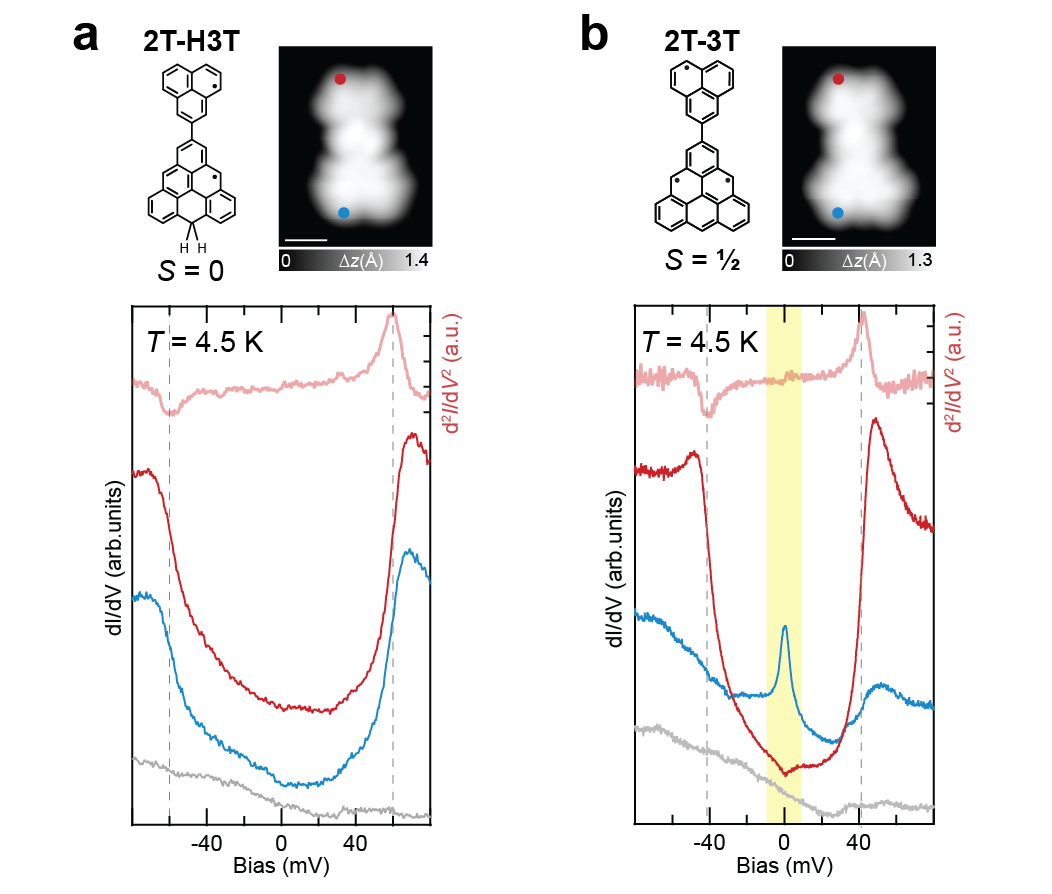}
    \caption{\textbf{Low-bias spectroscopy of 2T-H3T and 2T-3T.} Top (left to right): the chemical structure and corresponding STM image. \didv\ analysis of the dihydro compound 2T-H3T (a) and 2T-3T (b), acquired with the same CO-functionalized tip. The numerically computed \dsec, obtained after Gaussian smoothing of the \didv\ spectra, is shown in the top part of each graph. The 2T-3T species was obtained by tip-induced dehydrogenation of 2T-H3T. Open-feedback parameters: \mbox{--100~mV} / 850~pA; lock-in modulation \LI~=~1~mV. Scale bars: 0.5~nm.
}
    \label{fig:2T-H3T}
\end{figure}

To further demonstrate that the observed dip in the \didv\ signal is an intrinsic feature of the 2T-3T dimer—rather than an artifact induced by the tip---we performed a comparative high-resolution \didv\ analysis (at a temperature $T=4.5$ K ) using the same CO-functionalized tip. Initially, measurements were taken on the hydro intermediate 2T-H3T. We then homolytically cleaved the additional hydrogen atoms to obtain the 2T-3T structure and repeated the \didv\ analysis under identical conditions. As shown in Fig.~\ref{fig:2T-H3T}a, the spectra for 2T-H3T exhibit symmetric and intense inelastic excitations, previously attributed to singlet–triplet transitions in Ref.~\citenum{turco_arxiv}, with no sign of a zero-bias dip. Upon dehydrogenation to 2T-3T, however, distinct zero-bias peak and dip features emerge in the spectra. This direct comparison supports the conclusion that these spectroscopic features are intrinsic to the 2T-3T structure.

\subsection*{Analysis of the magnetic-field-dependent STS data}

For a spin-\nicefrac{1}{2} single-channel Kondo (1CK) system, applying a magnetic field $B_0$ perpendicular to the Au(111) surface is predicted to split the zero-bias resonance into two symmetric peaks once the Zeeman energy $g\mu_B B_0$ surpasses the Kondo singlet binding energy, characterized by a critical field $B_C \approx 0.5~{k_B T_K}/{g\mu_B}$ (for $T<0.25~T_K$)\cite{Costi:PRL:2000}. For $B < B_C$, the system remains in the strong Kondo coupling regime, and quasiparticle scattering is accurately described within Fermi liquid theory. Near the crossover region ($B \approx B_C$), the zero-bias resonance begins to split, and the residual strong exchange interaction between impurity and substrate electrons results in a renormalized g-factor\cite{Esat:PRR:2023,wolf_g-shifts_1969}. For $B > B_C$, the resonance fully splits into two symmetric peaks superimposed on conductance steps arising from inelastic spin-flip excitations; these features are well captured by a third-order perturbative scattering model in the exchange coupling\cite{ternes_probing_2017, Ternes:NJP:2015, Zhang:NComm:2013}. In this regime, both peak positions and excitation thresholds scale linearly with $B_0$, yielding a Landé $g$-factor of $g \approx 2$ \cite{mishra_topological_2020,li_uncovering_2020,Zhang:NComm:2013,Esat:PRR:2023}, consistent with the value expected for a free electron.

The zero-field differential conductance spectrum measured on 3T, in Fig. 3a, exhibits a narrow resonance accurately described by a Frota function (solid blue line) with a half width at half maximum (HWHM) of 54 µeV. Assuming standard Fermi-liquid behavior in the strong-coupling regime, this corresponds to a Kondo temperature of approximately $T_K \sim 140$ mK—two orders of magnitude lower than observed in similar spin-\nicefrac{1}{2} nanographenes on Au(111)\cite{Turco:PRR:2024,mishra_topological_2020}. Consistent with this low $T_K$, the resonance clearly splits into two symmetric peaks already at $B_0 = 1$ T, in agreement with the estimated critical field $B_C \sim 120$ mT (for $T>0.25~T_K$).

\subsection*{Strong Kondo coupling hypothesis}

\begin{figure}[ht!]
    \centering
    \includegraphics[width=\linewidth]{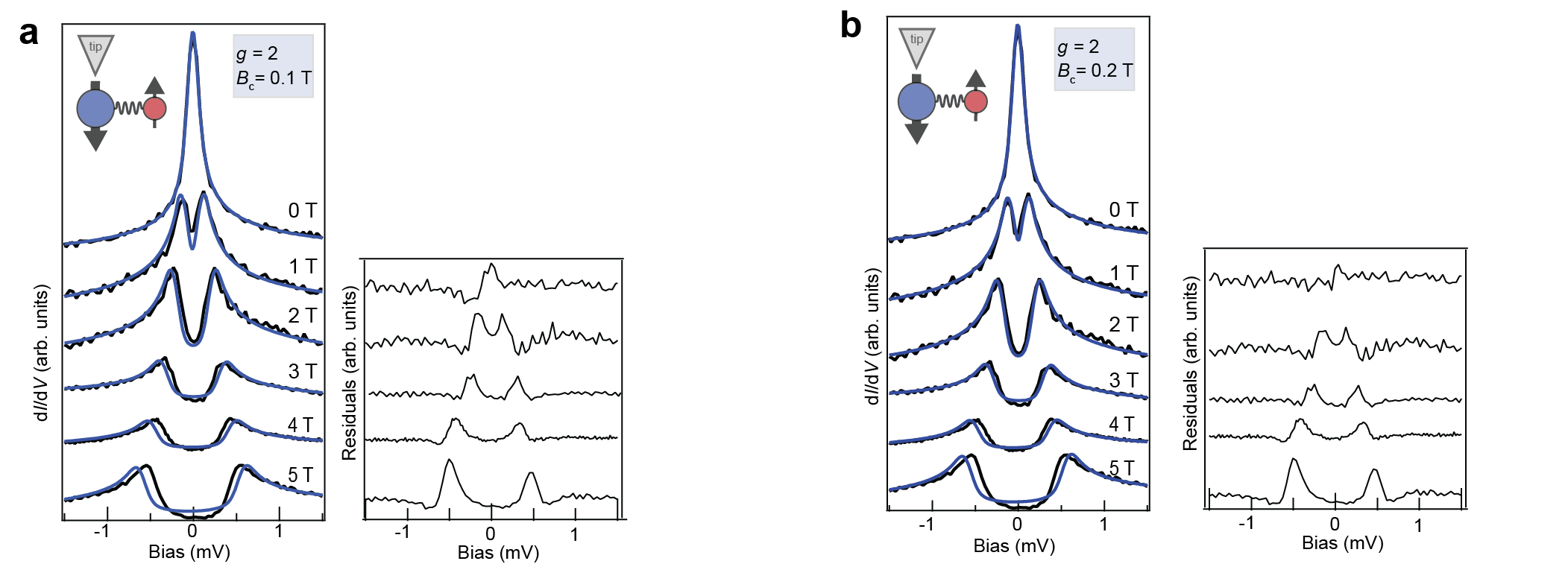}
    \caption{\textbf{A finite $B_{\text{C}}$ does not yield a good fit}. Experimental spectra (black) and fit (blue), as well as the residual (external panel) for $B_\text{C}=0.1$\,T (a) and $B_\text{C}=0.2$\,T (b).}
    \label{fig:failure finite Bc}
\end{figure}

Assuming the molecular spin behaves as a conventional spin-1/2 system in a magnetic field, we now show that this description fails to capture the true physical behavior. We therefore consider g-factor $g = 2$ and fit the $B$-field dependent spectra by varying $B_{\text{C}}$ as a fixed 
 input parameter. Convergence of the fit across all spectra was achieved only for $B_{\text{C}}< 0.2 $ T. Figure~\ref{fig:failure finite Bc} presents the fitting results with $g = 2$ and critical fields $B_{\text{C}}= 0.1 $ T and $B_{\text{C}}= 0.2 $ T. The residuals of each fit indicate that satisfactory fitting results cannot be obtained with $g = 2$ and  a fixed $B_{\text{C}}$.

 \subsection*{Weak Kondo coupling hypothesis}

\begin{figure}[ht!]
    \centering
    \includegraphics[width=0.7\linewidth]{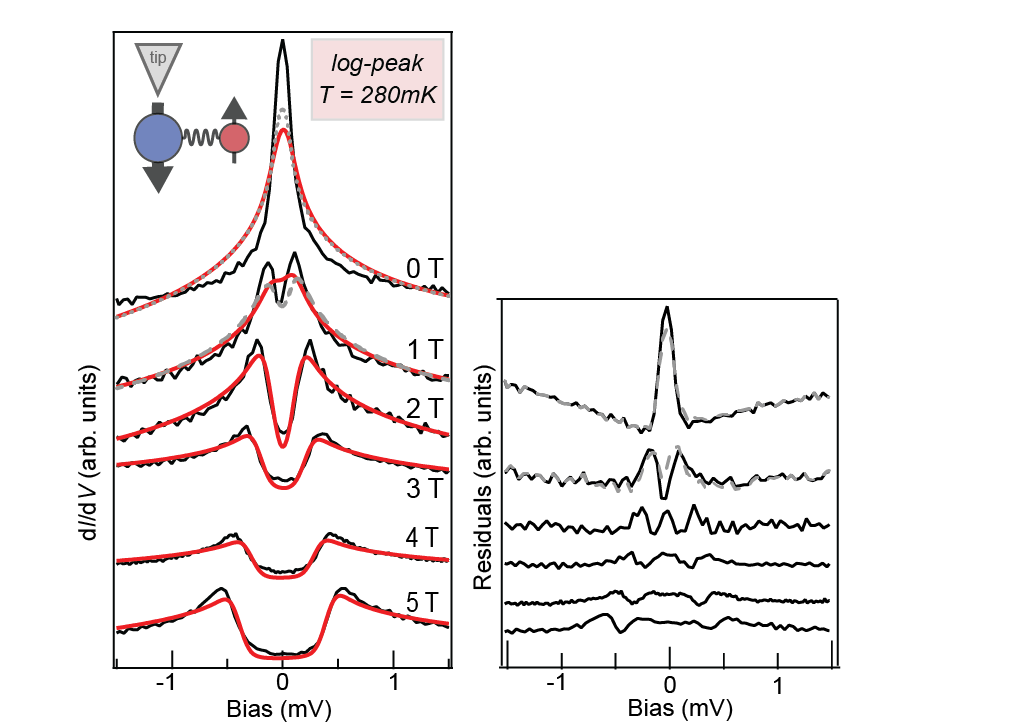}
    \caption{Attempted fit of split OKE peak (black) with a lock-in broadened ($80\,\mu\text{V}$) logarithmic peak (blue) at fixed temperature $T=280\,\text{mK}$ and $B_\text{eff}$ as fit parameter. The gray dotted line at $B=0\,\text{T}$ represents a fit using $T=60\,\text{mK}$. For the gray dashed line at $B=1\,\text{T}$ the effective B field was constrained to $B_\text{eff} > 0.45\,\text{T}$.
    All fits for finite B-field yielded $|J\rho| \gg 1$, indicating a contradiction with the weak coupling regime, which assumes $|J\rho| < 1$, in addition to the poor match between experiment and fitted spectra. Similarly, the residuals are plotted.
}
    \label{fig:logpeak}
\end{figure}

 For fields $B \geq 1$ T, the system enters the weak-coupling regime, where the peaks split by the magnetic field $B_0$ are theoretically expected to exhibit a logarithmic line shape\cite{Ternes:NJP:2015}. However, our analysis (see Fig. \ref{fig:logpeak}) reveal that the measured resonances cannot be adequately fitted by a logarithmic function. Instead, they are best captured by a temperature-broadened Frota spectral function, solid blue lines in Fig.~3a. Furthermore, the effective magnetic field $B$ extracted from these fits deviates significantly from the externally applied field $B_0$, as shown in Fig.~3b. This deviation indicates a renormalized Landé $g$-factor. 

 If we now instead set $B_{\text{C}}=0$ (and $g = 2$ ), we obtain the fitting results displayed in Fig. \ref{fig:g = 2, Bc = 0}, clearly demonstrating the impossibility to reproduce the experimental results.  

\begin{figure}[ht!]
    \centering
    \includegraphics[width=0.7\linewidth]{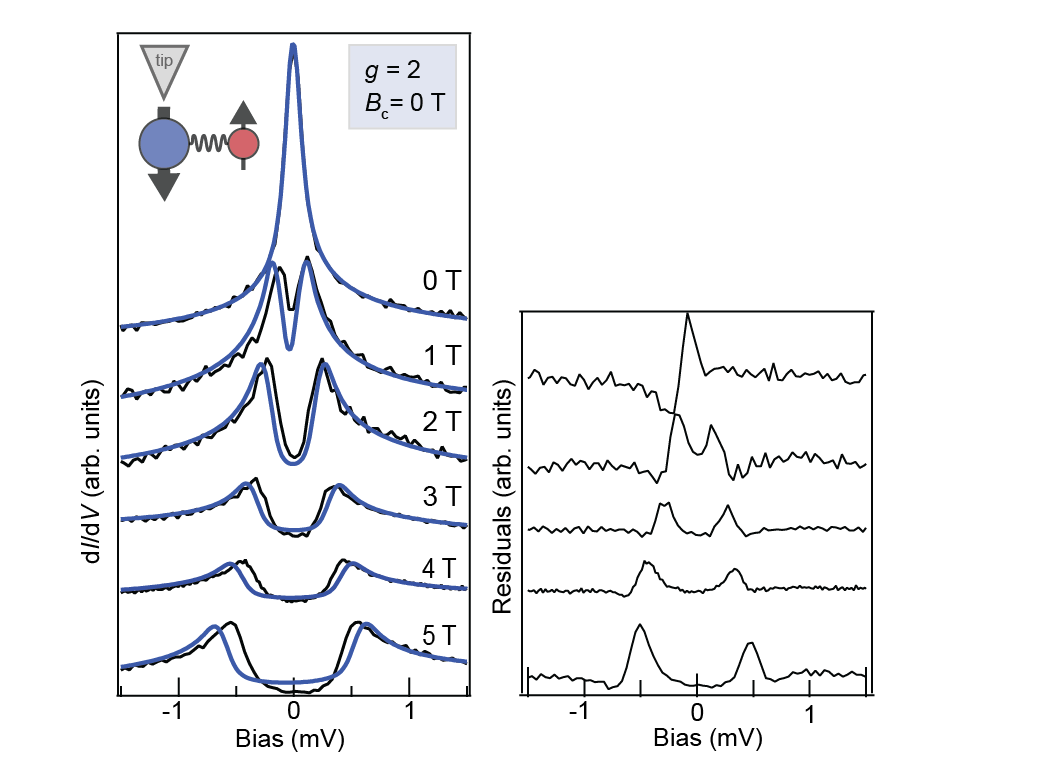}
    \caption{\textbf{Failure of fit with $g=2$ and $B_{\text{C}}=0$.} Experimental spectra (black) and fit (blue), as well as the residual (external panel).}
    \label{fig:g = 2, Bc = 0}
\end{figure}

\bibliography{refs}